\documentclass[usenatbib,useAMS]{mnras}
\usepackage{graphicx}
\usepackage{hyperref}
\usepackage[T1]{fontenc}
\usepackage{aecompl}

\makeatletter

\providecommand{\tabularnewline}{\\}

\usepackage{times}
\usepackage{amssymb}
\usepackage{color}

\newcommand{\kmps}{\mathrm{km~s^{-1}}}

\newcommand{\brg}{Br$\gamma$}
\newcommand{\pab}{Pa$\beta$}
\newcommand{\pad}{Pa$\delta$}
\newcommand{\ha}{H$\alpha$}
\newcommand{\hi}{H\,{\textsc{\lowercase{I}}}} 

\newcommand{\macc}{$\dot{M}_\mathrm{acc}$}
\newcommand{\lacc}{$L_\mathrm{acc}$}

\newcommand{\kms}{km\,s$^{-1}$}

\newcommand{\msyr}{M$_\mathrm{\odot}$\,yr$^{-1}$}

\newcommand{\lsun}{L$_\mathrm{\odot}$}

\newbox\grsign \setbox\grsign=\hbox{$>$} \newdimen\grdimen \grdimen=\ht\grsign
\newbox\simlessbox \newbox\simgreatbox
\setbox\simgreatbox=\hbox{\raise.5ex\hbox{$>$}\llap
     {\lower.5ex\hbox{$\sim$}}}\ht1=\grdimen\dp1=0pt
\setbox\simlessbox=\hbox{\raise.5ex\hbox{$<$}\llap
     {\lower.5ex\hbox{$\sim$}}}\ht2=\grdimen\dp2=0pt
\def\simgreat{\mathrel{\copy\simgreatbox}}

\title[VV\,Ser]{Investigating the origin and spectroscopic variability
  of the near-infrared H\,{\Large \textbf{I}} lines in the Herbig star VV\,Ser}
\author[R. Garcia Lopez et al.]{
Rebeca Garcia Lopez$^{1}$\thanks{E-mail:
rgarcia@cp.dias.ie (RGL); kurosawa@mpifr-bonn.mpg.de (RK)}, 
Ryuichi Kurosawa$^{2}$, 
Alessio Caratti o Garatti$^{1}$, 
Alexander Kreplin$^3$, \newauthor
Gerd Weigelt$^2$, 
Larisa V. Tambovtseva$^{2,4}$, 
Vladimir P. Grinin$^{2,4,5}$,
and Thomas P. Ray$^1$\thanks{Based on observations collected at the European Southern 
Observatory Paranal, Chile (ESO programme 093.C-0388(A)) and LBT, Mount Graham (AZ). The LBT is an international collaboration among institutions in the United States, Italy and Germany. 
LBT Corporation partners are: The University of Arizona on behalf of the Arizona Board of Regents; Istituto Nazionale di Astrofisica, Italy; LBT Beteiligungsgesellschaft, Germany, 
representing the Max-Planck Society, the Astrophysical Institute Potsdam, and Heidelberg University; The Ohio State University, and The Research Corporation, on behalf of The University of Notre Dame, University of Minnesota and University of Virginia.}
\\
$^{1}$ Dublin Institute for Advanced Studies, 31 Fitzwilliam Place, Dublin 2, Ireland \\
$^{2}$ Max-Planck-Institut f\"{u}r Radioastronomie, Auf dem H\"{u}gel 69, D-53121 Bonn, Germany \\
$^{3}$ AI School of Physics, University of Exeter, Physics Building, Stocker Road, Exeter EX4 4QL, UK \\
$^{4}$ Pulkovo Astronomical Observatory of the Russian Academy of Sciences, Pulkovskoe shosse 65, 196140, St. Petersburg, Russia \\
$^{5}$ The V.V. Sobolev Astronomical Institute of the St. Petersburg University, Petrodvorets, 198904 St. Petersburg, Russia \\
}
\begin{document}

\date{Accepted . Received  ; in original form }

\pagerange{\pageref{firstpage}--\pageref{lastpage}} \pubyear{2015}

\maketitle

\label{firstpage}

\begin{abstract}
The origin of the near-infrared (NIR) \hi\ emission lines in young stellar objects are not yet understood. To probe it, we present multi-epoch 
LBT-LUCIFER spectroscopic observations of the \pad, \pab, and \brg\ lines observed in the Herbig 
star VV\,Ser, along with VLTI-AMBER \brg\ spectro-interferometric observations at medium resolution.
Our spectroscopic observations show line profile variability in all the \hi\ lines. The strongest variability is observed in the redshifted part of the line profiles. 
The \brg\ spectro-interferometric observations  indicate that the \brg\ line emitting region is smaller than the continuum emitting region. To interpret our results, 
we employed radiative transfer models with three different flow configurations:  
magnetospheric accretion, a magneto-centrifugally driven disc wind, and a schematic bipolar outflow. Our models suggest that  the \hi\ line emission in 
VV\,Ser is dominated by the contribution of an extended wind, perhaps a bipolar outflow. 
Although the exact physical process for producing such outflow is not known, this model is capable of reproducing the averaged single-peaked line profiles of the 
\hi\ lines. Additionally, the observed visibilities, differential and closure phases are best 
reproduced when a wind is considered.
Nevertheless, the complex line profiles and variability could be explained by changes in the relative contribution of the magnetosphere and/or winds to the line emission.
This might indicate that the NIR \hi\ lines are formed in a complex inner disc region where inflow and outflow components might coexist. 
Furthermore,  the contribution of each of these mechanisms to the line appears time variable, suggesting a non-steady accretion/ejection flow.
\end{abstract}

\begin{keywords}
circumstellar matter -- infrared: stars.
\end{keywords}

\section{Introduction}

UX Orionis stars (UXORS) are pre-main sequence stars showing strong photometric variability at optical wavelengths. They receive their name after the prototype star UX Ori, a Herbig AeBe star. Most UXORS are, indeed, Herbig AeBe stars, although there are also
some UXORS amongst Classical T Tauri stars (CTTSs) of early K spectral type such as RY\,Tau and RY\, Lup. UXORS are characterised by sudden decreases in their optical brightness of up to 2-3\,mag. During the faint state, UXORS become redder with a 
 simultaneous increase of their polarisation. However, during deep minima these stars become bluer. % \citep{herbst99, natta00}. 
 %Most of UXORS sources are Herbig AeBe stars, but because of the unknown nature and origin of the UXOR phenomena they are still distinguished from the 
%classic Herbig AeBe definition. 
These observational properties, that is the magnitude fading, the increase of polarisation and the bluing effect, suggest that UXORS faiding is due to extinction events caused by filaments or clouds
of dust orbiting the central source and crossing the line of sight in a nearly edge-on disc \citep{grinin91,grinin98}. 
%Although this hypothesis reproduces most of the observational properties,  it has the caveat that 
%the dust responsible for the increase of extinction must be located in the outer disc region if a %flaring disc is considered, that is at distances scales of a hundred of au. Therefore, it would be %difficult to account for obscuration events lasting days or weeks 
%for this type of disc geometry.
A variation of this hypothesis was proposed by \cite{dullemond03} suggesting that UXORS events could only 
happen in self-shadowed disk geometries viewed at high inclination angles. In this model the inner disc is at the origin of the obscuring clouds, and it must have a puffed-up geometry.
In addition to the obscuration hypothesis, disc instabilities, inhomogeneities or warps can also contribute to the dimming magnitude observed in UXORS stars. For instance, \citep[][and references therein]{bouvier13} proposed a combination of these two scenarios 
to explain the complex photometric behaviour of AA\,Tau \citep[][and references therein]{bouvier13}. This source shows a low amplitude quasi-periodic variability that was explained by the presence of an asymmetric 
dusty warp at the disc inner edge caused by material lifted up from the disc plane by an inclined stellar magnetosphere. In addition to the low amplitude variability, AA\,Tau also shows deep minima consistent with a sudden increase of the dust extinction as predicted by the obscuration
hypothesis.

%%%%%%%%%%%%%%%%%%%%%%%%%%%%%%%%%%%%%%%%%%%%%%%%%%%%%%%%%%%%%%%%%%%%%%%%%%%%%%%
\begin{table}%[h!]
\caption{Total exposure times of the LBT observations. }
\label{tab:exposure_times}
\centering
\vspace{0.1cm}
\begin{tabular}{@{}l|ccc}
\hline
UT Date & z & J & K \\
     & \multicolumn{3}{c}{(s)}\\
\hline
2012-04-23 & 512 & 240 & 1280 \\
2012-05-28 & 720 & 480 & 128 \\
2012-06-23 & 720 & 240 & 64 \\
\hline
\end{tabular}
\end{table}
%%%%%%%%%%%%%%%%%%%%%%%%%%%%%%%%%%%%%%%%%%%%%%%%%%%%%%%%%%%%%%%%%%%%%%%%%%%%%%%%%

Despite many photometric studies, very little spectroscopic monitoring of UXORS has been conducted. The few studied cases show dramatic changes in the spectroscopic line profiles, not always linked with photometric variability \citep{eaton99,dewinter99,natta00_line,rodgers02}. 
In addition, most of these studies investigate variability in the optical, usually focusing on the H$\alpha$ line. 

In this paper we present a NIR spectroscopic monitoring of the UXOR Herbig AeBe star VV\,Ser. This source is located in the  main Serpens molecular cloud at $\sim$415\,pc \citep{dzib10}. 
There is some discrepancy in the literature about the photospheric temperature of VV\,Ser, but most studies identify VV\,Ser as an Herbig star of spectral type B. 
\cite{montesinos09} performed a complete 
study of the stellar properties of VV\,Ser, concluding that the most plausible temperature and age of this source is 13\,800\,K and 1.2\,Myr, respectively.
%
%A similar behaviour it is also observed in VV\,Ser an Herbig AeBe star located in the main Serpens molecular cloud at $\sim$415\,pc \citep{dzib10} and showing UXOR type variability. 
The light curve of VV\,Ser consists
of small amplitude brightness variations around the bright state
followed by rare non-periodical deep minima lasting of order of ten
days \citep{rostopchina01}.
The disc of VV\,Ser  appears to be self-shadowed, of low mass, and viewed at a high inclination angle \citep[i$\sim$70\degr;][]{pontoppidan07}. 
This would favour the obscuration hypothesis proposed by \cite{dullemond03} to explain the deep minima observed in VV\,Ser. 
%There is some discrepancy in the literature about the temperature of VV\,Ser, but most studies identify VV\,Ser as an Herbig star of spectral type B. 
%\cite{montesinos09} performed a complete 
%study of the stellar properties of VV\,Ser, concluding that the most plausible temperature and age of this source is 13\,800\,K and 1.2\,Myr.
In addition to the photometric variability typical of UXORS, VV\,Ser also shows spectroscopic line variability. In particular, optical spectroscopic studies show strong variability of the H$\alpha$ line profile \citep[e.g., ][]{mendigutia11}.  
To further constrain the structure of the inner gasseous disc of this source, additional medium-resolution interferometric observations are presented (Section\,\ref{sect:interferometric_obs}). Previous spectro-interferometric observations of this source show a compact \brg\ emitting region, tracing gas located 
within the dust sublimation radius \citep[][and references therein]{eisner14}.

%%%%%%%%%%%%%%%%%%%%%%%%%%%%%%%%%%%%%%%%%%%%%%%%%%%%%%%%%%%%%%%%%%%%%%%%%%%%%%%%%%%%%%%%%%%%%%%%%%%%%%%%%%%%%%%%%%%%%%%%%%%%%%%%%%%%%%%%%%%%%%%%%%%%%%%%%%%%%%%%%%%%%%%%%%%%%%%%%%
\begin{table*}%[h!]
\small
\begin{minipage}{140mm}
\caption{Observation log of the VLTI/AMBER medium-resolution (R$\sim$1500) observations of VV Ser conducted with the UT1-UT2-UT4 configuration. }
\label{obslog}
\centering
\vspace{0.1cm}
\begin{tabular}{@{}c c c c c c | c c }
\hline
UT Date & UT Time & DIT\footnote{Detector integration time per interferogram.} & NDIT\footnote{Number of interferograms.} & 
Proj. baselines & PA\footnote{Baseline position angle.} & Calibrator & UD diameter\footnote{The calibrator UD diameter (\textit{K} band) was taken from \cite{pasinetti01}.}\\ 
%UT Date & UT Time & DIT\tablefootmark{a} & NDIT\tablefootmark{b} &Proj. baselines & PA\tablefootmark{c} & Calibrator & UD diameter\tablefootmark{d}\\ 
 & [h:m] & [ms] & \# &  [m] & [$\circ$] & &[mas]\\  \hline 
2014-05-13 & 05:47--06:00 & 500 & 1200  &  47.66 / 76.34 / 104.98 & -165.47 / -99.14 / -123.72  & HD 163\,272 & $0.265 \pm 0.020$\\ 
2012-05-13 & 06:27--06:40 & 500 & 1200 &  49.20 / 83.25 /  115.28 & -156.67 / -98.40 / -120.38 &  HD 170\,920 & $0.285 \pm 0.019$\\ \hline
\end{tabular}

\end{minipage}
%\tablefoot{
%\tablefoottext{a}{Detector integration time per interferogram.}
%\tablefoottext{b}{Number of interferograms.}
%\tablefoottext{c}{Baseline position angle.}
%\tablefoottext{d}{The calibrator UD diameter (\textit{K} band) was taken from \cite{2001A&A...367..521P}.}
%}
\end{table*}
%%%%%%%%%%%%%%%%%%%%%%%%%%%%%%%%%%%%%%%%%%%%%%%%%%%%%%%%%%%%%%%%%%%%%%%%%%%%%%%%%%%%%%%%%%%%%%%%%%%%%%%%%%%%%%%%%%%%%%%%%%%%%%%%%%%%%%%%%%%%%%%%%%%%%%%%%%%%%%%%%%%%%%%%%%%%%%%%%%%%%%%%

This paper is structured as follows: The NIR spectroscopic and inteferometric observations along with the data reduction are presented in  Sections\,\ref{sect:spectroscopic_obs} and \ref{sect:interferometric_obs}. The results obtained from our observations
are shown in Sections\,\ref{sect:spectroscopy} (spectroscopy) and \ref{sect:interferometry} (interferometry). The theoretical models employed to interpret our results can be found in Section\,\ref{sect:models}, followed by a discussion (Section\,\ref{sect:discussion}) 
and the conclusions (Section\,\ref{sect:conclusions}).

%*******************************************************************
\begin{figure}
  \begin{center}
    \includegraphics[width=70mm]{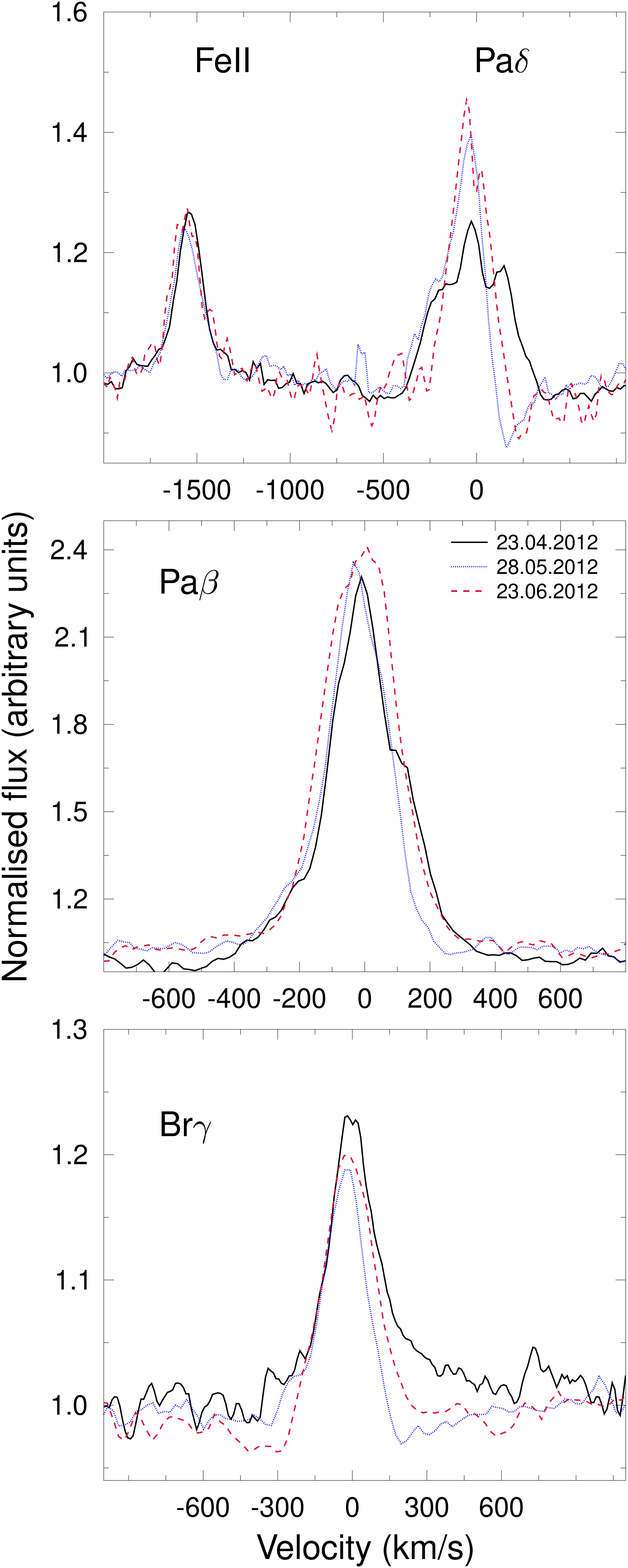}
  \end{center}
  \caption{LBT MR spectroscopic observations of VV\,Ser. The \hi\ line profiles are normalised to the continuum. The reported velocity is with respect to the local cloud velocity which is at an LSR velocity of +8\,\kms\ \citep{burleigh13}.}
  \label{fig:LBT-obs}
\end{figure}
%****************************************************************

\section{Observations and data reduction}
\label{sect:observations}

\subsection{Spectroscopic observations at medium resolution}
\label{sect:spectroscopic_obs}

VV\,Ser was observed during three different nights in 2012 (23 April, 28 May and 23 June) using the infrared spectrograph and camera LUCIFER, installed on the 
Large Binocular Telescope (LBT) at Mount Graham Observatory.  For each night, medium-resolution (MR) spectra in the z-, J-, and 
K-band were acquired consecutively, using the N1.8 camera (grating 210\_zJHK) and a slit width of 0\farcs5. This provides a spectral resolution of R$\sim$6700. The observations were performed using a spatial scale of 0.25\arcsec /pixel. 
Total exposure times for each date and band are listed in Table\,\ref{tab:exposure_times}. 

Data reduction was performed using standard IRAF tasks. Atmospheric OH lines were used for wavelength calibration. The atmospheric spectral response was corrected by dividing the object spectra
by the spectrum of a telluric spectroscopic standard. With this aim, the intrinsic absorption features of the standard were removed before dividing the spectra by a normalised blackbody at the appropriate 
temperature.
Finally, due to unstable seeing conditions as well as the presence of cirrus, no photometric calibration of the spectra was performed. 

\subsection{Near-IR interferometric observations}
\label{sect:interferometric_obs}

%\section{Observations and data reduction}
We observed VV Ser with the AMBER instrument of the Very Large Telescope Interferometer (VLTI) using the 8-m unit telescopes (UT's) with the configuration UT1-UT2-UT4 during one night in May 2014. 
The AMBER instrument is a three beam combiner that records spectrally dispersed three-beam interferograms \citep{petrov07}. 
Due to the faint H-band magnitude of VV\,Ser (H$\sim$7.4) and  poor atmospheric conditions (seeing$\sim$1\arcsec), the observations were conducted without the fringe tracker FINITO. We recorded 
\textit{K}-band interferograms, within the Br$\gamma$ line and the nearby continuum, in the medium spectral resolution mode (R$\sim$1500) with a detector integration 
time (DIT) of 500\,ms. To calibrate the atmospheric transfer function, we observed the calibrator stars HD\,163272 and HD\,170920. For data reduction, we employed the software library 
amdlib v3.0.5\footnote{The AMBER reduction package \textit{amdlib} is available at: http://www.jmmc.fr/data\_processing\_amber.htm} \citep{tatulli07, chelli09}. 
We selected 20\,per~cent of the frames with the highest 
fringe SNR as described in  \cite{tatulli07}. The observation log is shown in Table\,\ref{obslog}. 

\section{Spectroscopic results}
\label{sect:spectroscopy}

The most prominent features in our VV\,Ser spectra are hydrogen recombination emission lines, namely the Pa$\delta$ and Pa$\beta$ lines in the z- and J-band,  and the Br$\gamma$ line in the K-band. In addition, 
strong emission from the fluorescent Fe\,{\textsc{\lowercase{II}}}\,10000.3\AA\ line is also observed in the z-band \citep{bautista04,johansson04, ale13}. Figure\,\ref{fig:LBT-obs} shows the spectra of these lines, normalised to the continuum and calibrated in 
velocity for each of the three 
observation nights. 
All radial velocities are with respect to the cloud which is at a velocity of +8\,\kms\ with respect to the local standard of rest \citep[LSR;][]{burleigh13}. The measured radial velocities and equivalent widths (EWs) of each
line at each epoch are listed in Table\,\ref{tab:velocities}. The errors in the radial velocity values were estimated by comparing the theoretical wavelength of the OH lines located close to the line of 
interest in the sky frames and the measured value after wavelength calibration. 
% As reported in Table\,\ref{tab:velocities} and in 
%Figure\,\ref{fig:LBT-obs} the radial velocities and line profiles of all the detected recombination lines are different in each epoch. 

The  Fe\,{\textsc{\lowercase{II}}} line is roughly centred at zero velocity and does not show any significant line profile
variability. On the other hand, all the hydrogen recombination lines are, on average, blueshifted with EWs ranging from $\sim-$3\,\AA, for the \brg\ line, to $\sim-$17\,\AA, for the \pab. In addition, a night-by-night variation in the line EWs and line profiles is observed.

%In particular, line profile variability is observed in all the HI recombination lines (see, Fig.\,\ref{fig:LBT-obs}). 
In particular, the \pad\ line shows a very complex line profile with a clear line profile variability. 
In the first epoch (black continuous line in Fig.\,\ref{fig:LBT-obs}), the \pad\ line shows three different velocity components, namely a blueshifted high velocity component (HVC) at $\sim$--173\,\kms, a 
blueshifted low velocity component (LVC) at $\sim$--30\,\kms, and a redshifted velocity component at $\sim$+120\,\kms. In the second and third epoch (dotted blue and dashed red lines), 
the redshifted emission disappears as well as the blue-shifted double peak, leaving in its place an inverse P-Cygni profile.
Similar redshifted emission components are also observed during the first epoch in the \pab\ and \brg\ lines as a redshifted bump in the line profiles. As for the \pad\ line profile, these 
redshifted bumps disappear during the second and third epochs. In contrast with the \pad\ line, the redshifted absorption component of the inverse P-Cygni profile is only marginally observed in the \brg\ line during the last epoch.

%%%%%%%%%%%%%%%%%%%%%%%%%%%%%%%%%%%%%%%%%%%%%%%%%%%%%%%%%%%%%% %
\begin{table*}%[h!]
\begin{minipage}{140mm}
\caption{Radial velocities.}
\label{tab:velocities}
\centering
\vspace{0.1cm}
\renewcommand*{\thefootnote}{\alph{footnote}}
\begin{tabular}{@{}l|cc|cccc}
\hline
           & \multicolumn{2}{c}{2012-04-23} & \multicolumn{2}{c}{2012-05-28} & \multicolumn{2}{c}{2012-06-23} \\
Line id.   & V$_r$\footnotemark[1]
                            & EW\footnotemark[2]
                            & V$_r$\footnotemark[1]           & EW\footnotemark[2]           & V$_r$\footnotemark[1]    & EW\footnotemark[2]           \\
           & \kms\ & \AA & \kms\ & \AA & \kms\ & \AA \\                            
\hline
%FeII       & 4.7$\pm$3.2    & -57.0$\pm$19.7    & -6.0$\pm$6.5   &  -53.06$\pm$10.7        & -12.8$\pm$8.8   &  -55.7$\pm$14.0        \\    
FeII       & 4.7$\pm$3.2    & -1.9$\pm$0.6    & -6.0$\pm$6.5   &  -1.8$\pm$0.3        & -12.8$\pm$8.8   &  -1.9$\pm$0.5        \\ 
   %        & -208.4$\pm$3.2  & -23.3$\pm$8.3    &                &          &   &         \\
%Pa$\delta$ & -29.5$\pm$3.2    & -70.7$\pm$6.7    & -73.0$\pm$6.5  & -94.8$\pm$21.7              & -42.8$\pm$8.8               &   -110.3$\pm$25.7              \\
Pa$\delta$ & -29.5$\pm$3.2    & -2.4$\pm$0.2    & -73.0$\pm$6.5  & -3.2$\pm$0.7              & -42.8$\pm$8.8               &   -3.7$\pm$0.9              \\
           % & 119.6$\pm$3.2   & -55.8$\pm$10.2    &                &               &                &                \\
            & 119.6$\pm$3.2   & -1.9$\pm$0.3    &                &               &                &                \\
%Pa$\beta$  & -0.9$\pm$5.1    & -370.6$\pm$23.4  & -25.6$\pm$5.3  & -303.0$\pm$17.3        & -13.8$\pm$6.6  & -390.4$\pm$32.3         \\
Pa$\beta$  & -0.9$\pm$5.1    & -15.8$\pm$1.0  & -25.6$\pm$5.3  & -13.0$\pm$0.7        & -13.8$\pm$6.6  & -16.7$\pm$1.4         \\
Br$\gamma$ & 6.3$\pm$1.6     & -4.1$\pm$0.5    & -33.5$\pm$2.3  & -3.0$\pm$0.5         & -28.4$\pm$1.0  & -4.2$\pm$1.2          \\
\hline
\end{tabular}
\footnotetext[1]{Radial velocities are with respect to the LSR and corrected for a cloud velocity of 8\,\kms\ \citep{burleigh13}.}
\footnotetext[2]{Negative values of the equivalent width indicate the line is in emission.}
\end{minipage}
\end{table*}
%************************************************************

On the other hand, the radial peak velocities increase from the first to the second epoch in all the \hi\ lines, and they decrease from the second to the third epoch (see Table\,\ref{tab:velocities}).  
The lowest radial velocities are observed during the first epoch. This is due to the redshifted bumps observed during this epoch that shift the centred of the line profile towards the line rest velocity.
%(probably indicating that the two velocity components are not 
%spectrally resolved?). 
Almost the opposite behaviour is observed during the second epoch. 

%************************************************************
\begin{figure*}
\begin{center}
\includegraphics[width=175mm]{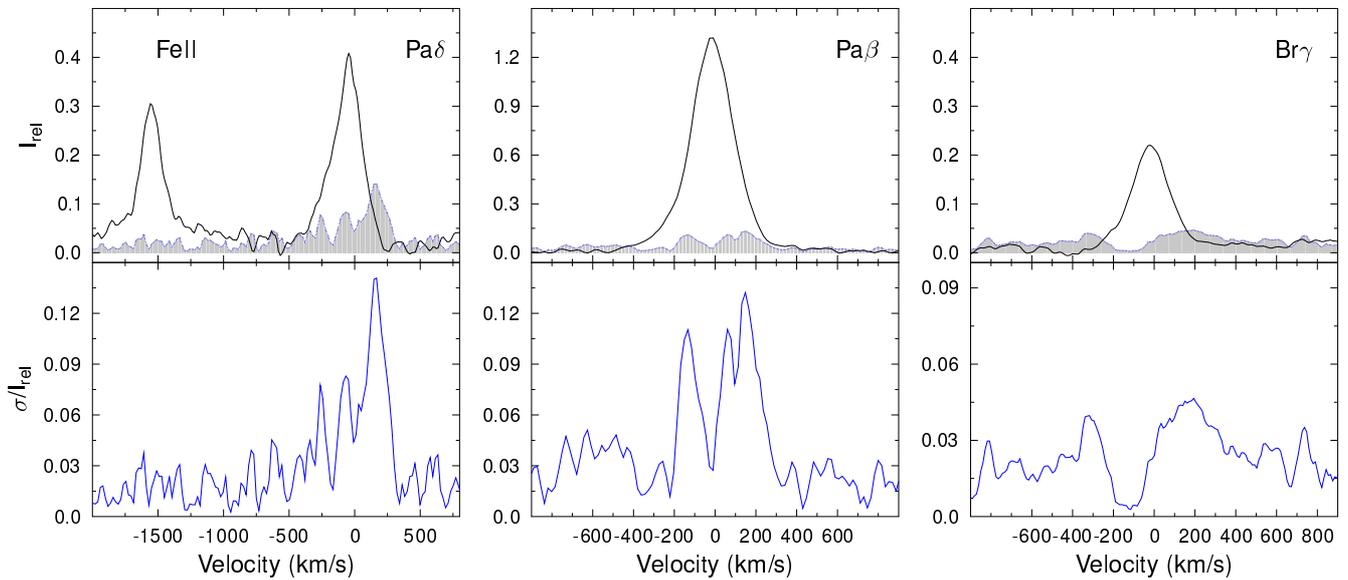}
\end{center}
\caption{{\bf Top:} From left to right, averaged line profile (I$_{rel}$) of the Fe\,{\textsc{\lowercase{II}}}\,10000.3\AA\ and \pad\ lines, the \pab\ line, and \brg\ line. The line profiles are normalised to the continuum, and shifted by -1. 
The  normalised variance profiles as in \textit{bottom} are overplotted (blue lines filled in grey) for comparison. {\bf Bottom:} Same as in \textit{Top}, but for the normalised variance line profiles ($\sigma$/I$_\mathrm{rel}$) of the Fe\,{\textsc{\lowercase{II}}} and \pad\ lines, 
the \pab\ line, and the \brg\ line, respectively.}
\label{fig:variance_profile}
\end{figure*}
%************************************************************

Similarly, EW variations are observed in the hydrogen recombination lines (see, Table\,\ref{tab:velocities}). These changes give us relative information on the continuum to line flux variations, indicating possible changes in the line or continuum fluxes through the analysed period. 
Unfortunately, the lack of simultaneous NIR photometry prevents us from distinguishing between both possibilities.
The measured EWs and errors are reported in Table\,\ref{tab:velocities}. %\footnote{The EW error was estimated from:
%$ \sigma(EW_\lambda) = \sqrt{1+\frac{F_c}{F}} \frac{(\Delta\lambda-EW_\lambda)}{S/N}$.
%Assuming a Poisson-statistics for the computation of the flux error and standard error propagation \citep{vollmann06}. 
%In this expression, $F_c$ and $F$ denote the continuum and line flux, respectively. 
%} 
%are reported in Table\,\ref{tab:velocities}. 
The EW error was estimated from:
\begin{equation}
 \sigma(EW_\lambda) = \sqrt{1+\frac{F_c}{F}} \frac{(\Delta\lambda-EW_\lambda)}{S/N}
\end{equation}
assuming Poisson-statistics for the computation of the flux error and standard error propagation \citep{vollmann06}. 
In this expression, $F_c$ and $F$ are the continuum and line flux, S/N is the 
signal-to-noise-ratio around the line position, and $\Delta\lambda$ is the wavelength interval 
over which the line EW was estimated. 
Both the \pab\ and \brg\ lines show night-to-night changes in their EWs. In particular, the EW decreases from the first to the second epoch in both lines, increasing again during the third epoch. Due to the complex \pad\ line profile,
it is difficult to extract any quantitative conclusion on the EW behaviour of this line. However, as mentioned above this line shows a clear line profile variability. 

To better quantify the line variability, the averaged and normalised variance profiles of the \pad, \pab, and \brg\ lines for the three epochs were computed (see, Fig.\,\ref{fig:variance_profile}). Following \cite{johns95}, the normalised variance profiles 
($\sigma_k$/I$_{\mathrm{rel},k}$) were derived by dividing the profile variance by the average line profile. 
The line profile variance was computed from:
\begin{equation}
\sigma_k = \Big[\frac{1}{(N-1)} \sum_{i=1}^{N} (I_{i,k}-I_{\mathrm{rel},k})^2\Big]^{1/2},
\end{equation}
where $I_{i,k}$ denotes the intensity profile of 
the line $k$ at each single epoch, I$_{\mathrm{rel},k}$ is the average profile
of the line $k$, and $N$ is the number of epochs.   The comparison of the averaged line profiles with the normalised 
variance profiles illustrates where the major changes in the line profiles take place \citep{johns95, kurosawa05, mendigutia11}. The most relevant changes across the line profiles affect the redshifted component of the line profiles, although 
smaller changes are also observed at blueshifted velocities. 
%component. 
Notably, both \pad\ and \pab\ lines show similar normalised variance profiles, indicating that their variability originates from similar physical processes. However, the line to continuum ratio of the \pad\ line is smaller than that for \pab, and therefore, 
the line profile variation is more evident in the former line.
The sizes of the normalised variances are slightly smaller for the \brg\ line compared to those of the \pab\ and \pad\ lines. As the \brg\ line is optically 
thinner than the \pab\ and \pad\ lines, the line emitting volume for these latter would be larger than that of the \brg. This might be the reason for the slightly
larger variability observed in the Paschen lines.

\subsection{Accretion properties}
\label{sub:acc-prop}

The accretion luminosity of VV\,Ser was estimated from the luminosity of the \brg\ line using the empirical relationship derived from \cite{donehew11}. The \brg\ luminosity was derived from the EW of the line, once the 
intrinsic photospheric absorption contribution to the line was corrected \citep[see][for more details]{rebeca06, donehew11}. This leads to \brg\ photospheric-corrected equivalent widths (EW$_\mathrm{circ}$) of 4.7, 3.6 and 4.8\,\AA\ for the first, 
second and third epoch, respectively. The stellar parameters used in the computation can be found in Table\,\ref{tab:stellar_parameters}. In addition, a visual extinction of A$_V$=3.4\,mag was assumed \citep{montesinos09}. 
The accretion luminosity (\lacc) was estimated  assuming a constant 2MASS K-band magnitude of the source of m$_K$=6.32. The  average \lacc\ value over the three epochs is $\sim$15.8$^{+9.3}_{-7.9}$\,\lsun. 
This leads to a mass accretion rate of \macc $\sim$ 3.1$^{+1.6}_{-1.5}\times$10$^{-7}$\,\msyr, assuming that \macc= \lacc\ R$_*$/ G M$_*$. In this expression, R$_*$ and M$_*$ are the stellar radius and mass (see Table\,\ref{tab:stellar_parameters}), and G is the universal gravitational constant. 
%This value is consistent with the one reported in \cite{donehew11} once the difference in the assumed distance is taken into 
%account.
%Note for us: using the EWcirc of Donehew = -10.8, and our distance, the value obtained is log Lacc=1.5, and Macc=7.1e-7(-3.6,+7.2).

%%%%%%%%%%%%%%%%%%%%%%%%%%%%%%%%%%%%%%%%%%%%%%%%%%%%%%%%%%%%%%%%%%%%%%%%%%%%%%%%%%%%%%%%%%%%%%%%%%%%%%%%%%%%%%%%%%%%%%%%%%%%%%%%%
\begin{table*}
\begin{minipage}{140mm}
\caption{Geometric models}
\label{tab:geometric_model} 
\begin{center}
\renewcommand*{\thefootnote}{\alph{footnote}}
\begin{tabular}{c | ccc | ccc}
\hline 
& \multicolumn{3}{c}{Continuum} & \multicolumn{3}{c}{Line} \\
   & R & PA & $i$ & $R$ & PA & $i$\\
   & (mas) & \multicolumn{2}{c}{(deg)} & (mas) &  \multicolumn{2}{c}{(deg)} \\       
   \hline
Ring fitting & 2.3$\pm$0.1 & 66.7$\pm$1.6 & 70.2$\pm$1.8 & 2.3$\pm$ 0.3 & 61.3$\pm$4.0 & 70.3$\pm$4.4 \\
Gaussian fitting & 4.0$\pm$0.2 & 68.4$\pm$2.3 & 61.6$\pm$2.1 & 4.1$\pm$0.6 & 61.0$\pm$4.7 & 70.4$\pm$3.1 \\
\hline 
\end{tabular}
\par\end{center}
\end{minipage}
\end{table*}
%%%%%%%%%%%%%%%%%%%%%%%%%%%%%%%%%%%%%%%%%%%%%%%%%%%%%%%%%%%%%%%%%%%%%%%%%%%%%%%%%%%%%%%%%%%%%%%%%%%%%%%%%%%%%%%%%%%%%%%%%%%%%%%%%%%%

\section{Interferometric results}
\label{sect:interferometry}

\begin{figure}
  \begin{center}
    \includegraphics[width=0.3\textwidth]{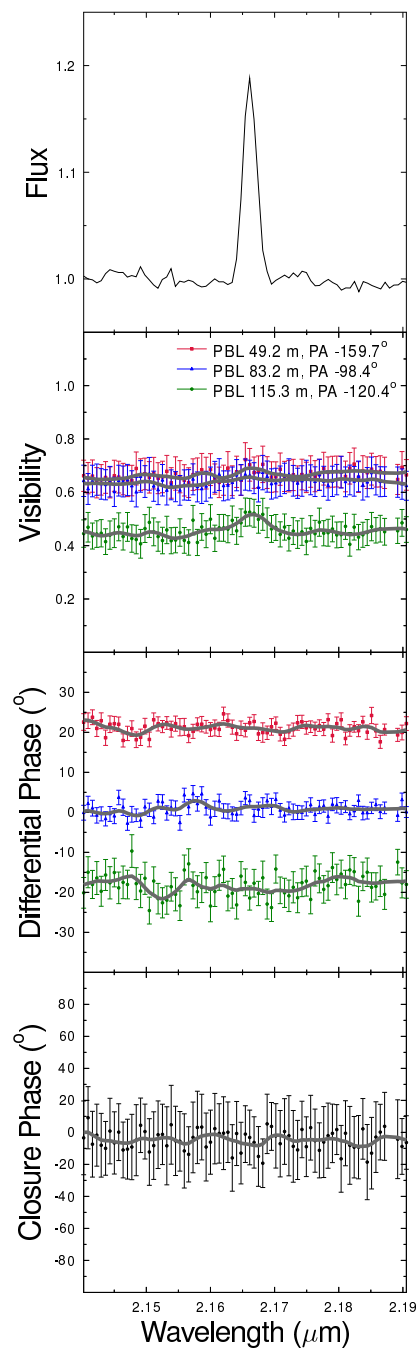}%_crop}
  \end{center}
  \caption{AMBER observation of VV\,Ser with three different baselines and spectral resolution R=1500. From top to bottom: \brg\ line profile (normalised to the continuum), visibilities, wavelength-dependent differential phases, and closure phase. A spectral binning to the raw data (from R=1500 to R=750) is overplotted in grey (solid line).  
  For clarity, the differential phases of the shortest and longest baselines have been shifted by +20$^o$ and -20$^o$.}
  \label{fig:AMBER-obs}
\end{figure}

Our interferometric observations provide four direct observables: the \brg\ line profile, visibilities, differential phases and closure phase.
These observables give us information about the size and kinematics of the \brg\ emitting region.
Figure\,\ref{fig:AMBER-obs} shows the results from our AMBER-MR observations of VV\,Ser. From top to bottom the \brg\ line profile, three baseline visibilities, differential phases and closure phase are shown. The \brg\ line profile is very similar to the one detected in our 
LUCIFER-LBT observations, that is, 
 the line profile is single-peaked with a triangular shape. The wavelength dependent visibility curve (second panel) is quite flat, indicating that the \brg\ line is emitted in a 
region of similar size to the continuum emitting region. At the longer baseline, a faint increase of the visibilities within the \brg\ line is detected. This would indicate that most
of the \brg\ emission is produced in a region smaller than the dust sublimation radius, as previously observed in other Herbig AeBe stars \citep[e.g.,][]{kraus08,rebeca15, ale15}.

By fitting an elongated Gaussian model to the visibilities shown in Fig.\,\ref{fig:AMBER-obs}, we  estimated a half width at half maximum (HWHM) for the  continuum emitting region of 4.0$\pm$0.2\,mas (1.65$\pm$0.08\,au) with an elongation ratio of $\sim$2.7, corresponding to an 
inclination angle ($i$) of 61\fdg6$\pm$2\fdg1. The positon angle (PA) of the system axis is 68\fdg4$\pm$2\fdg3. 
Alternatively, if an inclined ring model with a ring width of 20\,per~cent of the inner ring radius is used, then an inner radius of 2.3$\pm$0.1\,mas (0.97$\pm$0.01\,au) is found. This corresponds to an inclination angle of 70\fdg2$\pm$1\fdg8, and a PA of 66\fdg7$\pm$1\fdg6. 
Finally, none differential phases and closure phase signal are detected within the uncertainties of 5\degr and 20\degr, respectively. 

%*******************************************************************
%****************************************************************

%*******************************************************************
\begin{figure}
  \begin{center}
    \includegraphics[width=0.30\textwidth]{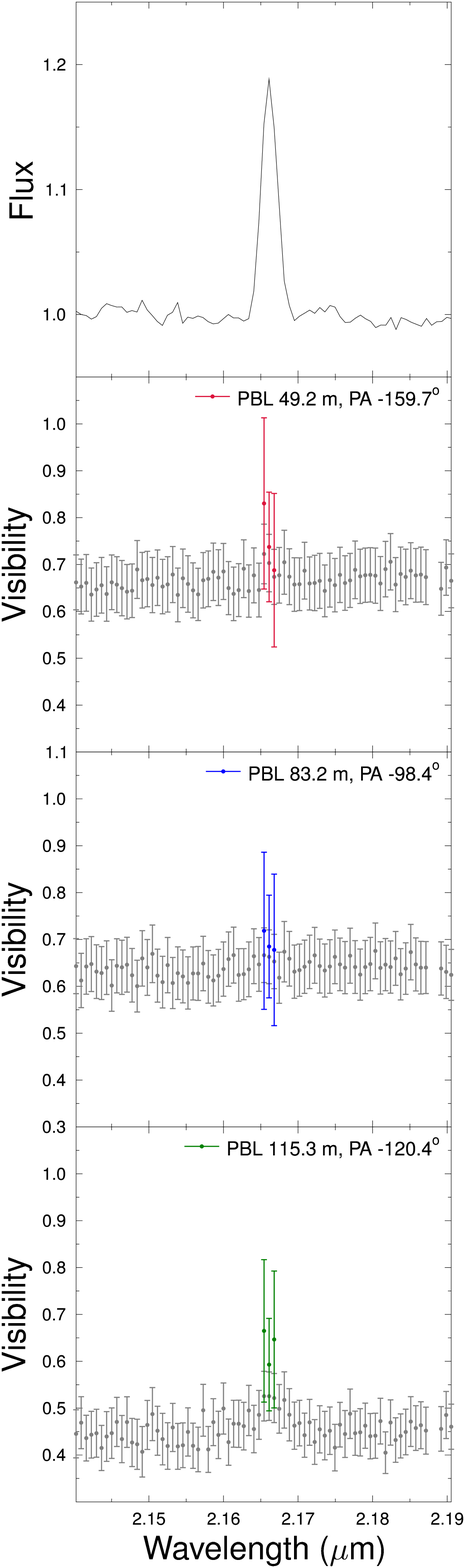}%_crop}
  \end{center}
  \caption{Observed AMBER pure \brg\ line visibilities of
    VV\,Ser. From top to bottom: wavelength dependence of flux, visibilities of the first, second, and third baseline. In each visibility panel the observed total visibilities (grey), and the observed continuum-compensated pure
\brg\ line visibilities of each baseline  (red, blue and green) are shown.}
  \label{fig:pure_line_vis}
\end{figure}
%****************************************************************

\subsection{The size of the Br$\gamma$ emitting region}
\label{sect:geometric_model}

Our AMBER-MR observations of VV\,Ser allow us to measure the dispersed pure \brg\ line visibilities at three different radial velocities across the \brg\ line. To measure this continuum-compensated visibilities, we used the method described in \cite{weigelt07}. According to this technique, 
within the wavelength region of the \brg\ line, the visibility has two components: the pure-line emitting component and the continuum emitting component. This latter includes contributions from both the circumstellar environment and the stellar component. Therefore, the emission 
line visibility $V_{Br\gamma}$ in each spectral channel can be written as $ F_{{\rm Br}\gamma}V_{{\rm Br}\gamma} = |F_{\rm tot}V_{\rm tot} - F_{\rm c}V_{\rm c}| $ if the differential phase is zero \citep[see][for more details]{weigelt07}. The values $F_{\rm tot}$ and $V_{\rm tot}$ denote 
the total measured flux and visibilities in the \brg\ line, and $F_{\rm c}$ and $V_{\rm c}$ the continuum flux and visibility, respectively.
This procedure takes then into account the contribution of the intrinsic photospheric absorption feature of VV\,Ser by considering a synthetic spectrum of the same spectral type and surface gravity (see, Table\,\ref{tab:stellar_parameters}). 
The resulting pure \brg\ line visibilities are shown in Fig.\,\ref{fig:pure_line_vis}. Only spectral channels with a line flux higher than 10\,per~cent of that of the continuum were considered.

The average line visibilities are 0.75, 0.70, and 0.65 for the shortest ($\sim$50\,m), medium ($\sim$83\,m) and longest ($\sim$115\,m) baseline, respectively. Similarly to the continuum fitting,  an approximate size of the \brg\ line emitting can be derived from the pure line visibility 
values. 
%The results are shown in Table\,\ref{tab:geometric_model}, however, due to the large errors, no difference between the continuum and \brg\ line emitting regions is found.

\section{Modelling spectroscopic and interferometric observations}
\label{sect:models}

To constrain the physical processes dominating VV\,Ser circumstellar environment, we model the emission line profiles obtained 
with LBT-LUCIFER (Section\,\ref{sect:spectroscopy}) and the VLTI-AMBER
interferometric data (e.g.~visibility and differential phases; Section\,\ref{sect:interferometry}).

To model the line profiles, we use the radiative transfer
code {\sc torus} (e.g.~\citealt{harries00,symington05, kurosawa06,kurosawa11}).  In particular, the numerical method used in the
current work is essentially identical to that in
\citet{kurosawa11}, originally developed to study CTTSs. The model employs the adaptive mesh refinement
grid in Cartesian coordinate and assumes an axisymmetry around the
stellar rotation axis. The model includes 20 energy levels of the hydrogen
atom. The non-local thermodynamic equilibrium (non-LTE) level
populations are computed using the Sobolev approximation
\citep[e.g.][]{sobolev57,castor70,castor79}.  For more
comprehensive descriptions of the code, readers are referred to
\citet{kurosawa11}.

For the work presented here, three minor modifications have been, however,
introduced in the code described in \citet{kurosawa11}. 
Firstly, the
model now includes continuum emission from a geometrically narrow
ring which emulates the K-band dust emission near the dust sublimation
radius.  This emission contribution is important for modelling a
correct line strength of the Br$\gamma$ line as well as the
interferometric quantities around the line. 

Secondly, the effect of a rotating
magnetosphere has been included using the method
described in \cite{muzerolle01}.  Until now, the code has been applied
only to CTTSs. These stars are slow rotators, with rotation periods of 2--10~d (e.g.~\citealt{herbst87};
\citealt{herbst94}). Hence, the effect of the rotation in CTTSs are small/negligible (e.g.~\citealt{muzerolle01}). Conversely, 
intermediate-mass Herbig Ae/Be stars are expected to be fast rotators with periods of $0.2$ to a few days
\citep[e.g.][]{hubrig09, hubrig11}. Therefore, the effect of rotation should be included for modelling VV\,Ser.

Thirdly, the code has been modified to write model intensity maps
as a function of wavelength (or velocity bins). This will allow us to
compute the interferometric quantities, such as visibility,
differential and closure phases, as a function of wavelength, which
can be then directly compared with the spectro-interferometric observations
of VLTI-AMBER. 

The difference in the rotation properties of CTTSs and Herbig AeBe stars could be due the weak magnetic fields detected in Herbig AeBe stars. 
Indeed, spectro-polarimetric surveys of Herbig AeBe stars recently revealed the presence of weak magnetic fields in these sources, although 
much weaker than those measured in CTTSs \citep[from few to several hundreds of Gauss vs. the some kG measured in CTTSs;][]{hubrig09, alecian13}. 
Additonal evidence of magnetic fields is given by the fact that many Herbig AeBe stars are also  X-ray sources \citep{hubrig09,drake14}.
As a consequence, the magnetic truncation 
radius of Herbig AeBe stars is expected to be located much closer to the 
protostar than in CTTSs \citep[see, e.g.][]{shu94,muzerolle04}, giving rise to very compact, fast rotating, magnetospheres. 
%Accordingly, Herbig AeBe should have much more compact magnetospheres than CTTSs. 
This scenario is additionally supported by recent NIR and optical spectroscopic surveys of Herbig AeBe stars
\citep[see, e.g.][]{cauley14,cauley15, fairlamb15}. In particular, these studies show clear evidences of inverse 
P-Cygni profiles in Herbig Ae and late type 
Herbig Be stars (indicative of accretion), as well as blue-shifted absorption features in strong accretors (indicative of accretion-driven outflows). 
%accreting objects usually display
%blue-shifted absorption features, suggesting that the outflows in Herbig AeBe stars are
%accretion-driven. 
%The maximum inflow speeds found in these sources are, however, smaller than those found in CTTSs, suggesting a more compact 
%magnetosphere in Herbig Ae and late type Herbig Be stars than in CTTSs.  
These findings, along with the lack of inverse P-Cygni profiles in Herbig Be stars, suggest
a general transition from magnetically controlled
accretion in Herbig Ae and late type Be stars to boundery layer accretion in Herbig Be stars \citep[see][]{cauley15,fairlamb15}.

For the particular case of VV\,Ser, attempts at measuring the magnetic field were firstly performed by \cite{hubrig09}. 
This study, was however only capable of diagnosing VV\,Ser's magnetic field at a significant level of 
2.7$\sigma$ (200$\pm$75\,G), and thus VV Ser was only classified as an Herbig star "suggestive" of
 the presence of a magnetic field. Additional attempts were performed by \cite{alecian13}. In this case, 
 a magnetic field
of 561\,G was measured, although with a large uncertainty of $\pm$238 G. The difficulty in measuring the 
magnetic field of VV\,Ser through spectro-polarimetry is probably due to the  combination of the
faintness of this source (V=11.6, and a nearly edge-on disc, self-shadowed by a puffed-up
inner disc rim, see \citealt{pontoppidan07}), along with the large measurement uncertainties of this technique.
However, VV\,Ser shows inverse P-Cygni profiles in several optical lines \citep[see][]{mendigutia11, cauley15}, as well
as redshifted absorption components in the Pa$\delta$ and Br$\gamma$ lines observed in our  LBT NIR spectra (see Fig.\,\ref{fig:LBT-obs}). 
For these reasons, there is the possibility that VV Ser is weakly magnetised and has a compact 
magnetosphere and an accretion driven wind. Therefore, in the following sections we explore this possibility, and further assume that the outflow
mechanisms in VV Ser are similar to those in the CTTSs which are also magnetised but
have stronger magnetic fields. In particular, we examine our spectroscopic and interferometric
data using combinations of a magnetosphere, a disc wind and a bipolar wind. 

Finally, it is worth noting that the advantages of the adopted model over a simple geometric model are several. Firstly, 
our modelling can provide us with estimates of basic physical parameters under our assumptions, such as the wind
  mass-loss rate, the gas temperature and density, in addition to the
  emission geometry. Secondly, it allows us to model the emission lines
  observed with LBT (\pad, \pab, and \brg) simultaneously, and in a self-consistent way.

\subsection{Model configurations }
\label{sub:model-config}

%The basic schematic diagrams of our models are shown in
Our model sketches are shown in Fig.~\ref{fig:model-config}.  The three models consider three distinct flow
components: (1)~a dipolar magnetospheric accretion zone (magnetosphere) as described in
\citet{hartmann94} and \citet{muzerolle01}, (2)~a disc wind
emerging from the equatorial plane (a geometrically thin
accretion disc) located outside of the magnetosphere (Fig.\,\ref{fig:model-config}, left) and
(3)~a bipolar outflow in the polar regions of the
  star (Fig.\,\ref{fig:model-config}, right). In the following, we briefly describe each flow component and
their key parameters. Detailed descriptions of each component can be
found in \citet{kurosawa11}.

\subsubsection{Magnetosphere }
\label{subsub:MA-config}

The accretion stream through a dipolar magnetic field is described
as $r=R_{\mathrm{m}}\,\sin^{2}{\theta}$ (e.g.~\citealt{ghosh77};
\citealt{ghosh79}; \citealt{hartmann94}) where $r$ and $\theta$
are the polar coordinates; $R_{\mathrm{m}}$ is the magnetospheric
radius at the equatorial plane. The accretion funnel regions are defined
by two stream lines corresponding to the inner and outer magnetospheric
radii, that is, $R_{\mathrm{m}}=R_{\mathrm{mi}}$ and $R_{\mathrm{mo}}$
(Fig.~\ref{fig:model-config}). We adopt the density and temperature
structures along the stream lines as in \cite{muzerolle01}.
The temperature scale is normalised with a parameter $T_{\mathrm{m}}$
which sets the maximum temperature in the stream. The mass-accretion
rate $\dot{M}_{\mathrm{a}}$\footnote{To avoid confusion, from now on $\dot{M}_{\mathrm{a}}$ refers to the input model parameter and \macc\ to the measured value.} scales the density of the magnetospheric
accretion funnels.

\subsubsection{Disc wind}
\label{subsub:DW-config}

Our disc wind model is an adaptation of the ``split-monopole'' wind
model by \citet{knigge95} and \citet{long02}. This simple
kinematic wind model roughly follows the basic ideas of the
magneto-centrifugal wind paradigm (e.g.~\citealt{blandford82}). The
outflow arises from the surface of the rotating accretion disc, and
has a biconical geometry. The specific angular momentum is assumed to
be conserved along stream lines, and the poloidal velocity component
is assumed to be simply radial from ``sources'' vertically displaced
from the central star. Near the disc surface where the wind emerges,
the value of the velocity toroidal component is similar to that of the local
Keplerian velocity of the disc.
%=========================================
\begin{table*}
\begin{minipage}{140mm}
\caption{Adopted parameters of the star and magnetosphere for VV~Ser}
\label{tab:stellar_parameters} 
\begin{center}
\renewcommand*{\thefootnote}{\alph{footnote}}
\begin{tabular}{cccccccccc}
\hline 
$R_{*}$  & $M_{*}$  & $T_{\mathrm{eff}}$  & $\log g$ & $P_{*}$ & $d$  & $\dot{M}_{\mathrm{a}}$ & $R_{\mathrm{mi}}$ & $R_{\mathrm{mo}}$ & $T_{\mathrm{m}}$\tabularnewline
($\mathrm{R_{\odot}}$)  & ($\mathrm{M_{\odot}}$)  & ($\mathrm{K}$)  & (cgs) & (d) & ($\mathrm{pc}$)  & ($\mathrm{M_{\odot}yr^{-1}}$) & ($R_{*}$) & ($R_{*}$) & ($\mathrm{K}$) \tabularnewline
\hline 
$2.6$\footnote{From \cite{montesinos09} assuming a distance of 415\,pc}  & $4.0$\footnotemark[1]  & $14000$\footnotemark[1]  & $4.0$\footnotemark[1] & $0.86$ & $415$\footnote{From \cite{dzib10}}  & $3.3\times10^{-7}$ & $1.3$ & $1.8$ & $10000$\tabularnewline
\hline 
\end{tabular}
\par\end{center}
\end{minipage}
\end{table*}
%=========================================

\begin{figure}
\begin{center}
\includegraphics[width=0.48\textwidth]{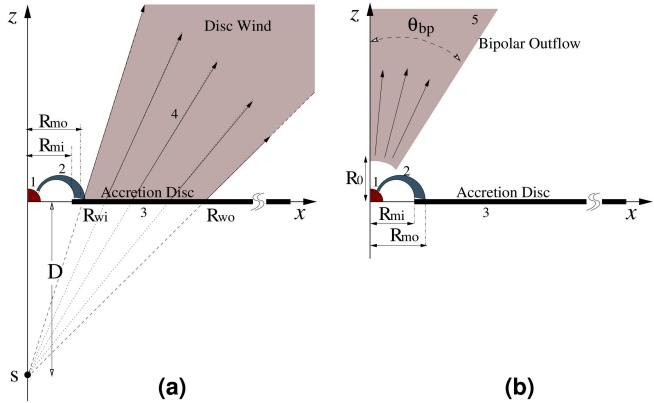}
\end{center}
\caption{Basic model configurations of the circumstellar flows which
  consist of (a)~a magnetosphere and a disc wind (left panel), or (b)~a
  magnetosphere and a bipolar outflow (right panel). The density is assumed
  to be axisymmetric and top-down symmetric. The components of the
  models are: (1)~the continuum source (star) located at the origin of
  the cartesian coordinates system, (2)~the magnetospheric accretion
  flow, (3)~the geometrically thin accretion disc, (4)~the disc wind,
  and (5)~the bipolar outflow. The split-monopole (with the source
  displacement distance $D$) disc wind emerges from the equatorial
  plane, but only within the distances from $z$-axis between
  $R_{\mathrm{wi}}$ and $R_{\mathrm{wo}}$. The bipolar outflow is
  launched from a sphere with radius $R_{\mathrm{0}}$, but is
  restricted within the cones with the half opening angle
  $\theta_{\mathrm{bp}}$. The figure is not to scale.}
\label{fig:model-config} 
\end{figure}

Our model has five basic parameters: (1)~the total mass-loss rate in the disc
wind ($\dot{M}_{\mathrm{dw}}$), (2)~the degree of the wind
collimation, (3)~the steepness of the radial velocity
($\beta_{\mathrm{dw}}$), (4)~the wind temperature, and (5)~the 
power-law index of the local mass-loss rate per unit area 
$\dot{m}(w) \propto w^{-p}$ where $w$ is the distance from the star 
on the equatorial plane.  The basic
configuration of the disc-wind model is shown in
Fig.~\ref{fig:model-config}. The disc wind originates from the disc
surface, and the ``source'' points ($S$), from which the stream lines
diverge, are placed at a distance $D$ above and below the centre of the
star. The angle at which the matter is launched from the disc is controlled
by changing the value of $D$. The mass-loss launching occurs between
$R_{\mathrm{wi}}$ and $R_{\mathrm{wo}}$ where the former is set to be
near the outer radius of the closed magnetosphere
($R_{\mathrm{mo}}$) and the latter to be a free parameter.  The wind
density is normalised with the total mass-loss rate in the disc wind
($\dot{M}_{\mathrm{dw}}$). The temperature of the wind
($T_{\mathrm{dw}}$) is assumed as isothermal.

%=========================================
\begin{table*}
\caption{Model parameters for disc wind and bipolar outflow.}
\label{tab:model-param}
\begin{tabular}{lcccccccccccccccc}
\hline 
 & MA$^{\dagger}$ &  & \multicolumn{7}{c}{Disc wind} &  &
\multicolumn{6}{c}{Bipolar outflow}\tabularnewline
\cline{2-2} \cline{4-10} \cline{12-17} 
 & (1) &  & (2) & (3) & (4) & (5) & (6) & (7) & (8) &  & (9) & (10) & (11) & (12) & (13) & (14)\tabularnewline
 &  &  & $\dot{M}_{\mathrm{dw}}$ & $R_{\mathrm{wi}}$ & $R_{\mathrm{wo}}$ & $p$ & $T_{\mathrm{dw}}$ & $\beta_{\mathrm{dw}}$ & $D$ &  & $\dot{M}_{\mathrm{bp}}$ & $\theta_{\mathrm{bp}}$ & $R_{0}$ & $T_{\mathrm{bp}}$ & $V_{\infty}$ & $\beta_{\mathrm{bp}}$\\
Model  &  &  & ($\mathrm{M_{\odot}yr^{-1}}$) & ($R_{*}$) & ($R_{*}$) & $\cdots$ & ($\mathrm{K}$) & $\cdots$ & ($R_{*}$) &  & ($\mathrm{M_{\odot}yr^{-1}}$) & $\cdots$ & ($R_{*}$) & ($\mathrm{K}$) & ($\mathrm{km\, s^{-1}}$) & $\cdots$\\
\hline 
A$^{*}$ & yes &  & $\cdots$ & $\cdots$ & $\cdots$ & $\cdots$ & $\cdots$ & $\cdots$ & $\cdots$ &  & $\cdots$ & $\cdots$ & $\cdots$ & $\cdots$ & $\cdots$ & $\cdots$\\
B & yes &  & $3\times10^{-8}$ & $1.8$ & $10.0$ & $2.3$ & $10^{4}$ & $4.0$ & $2.2$ &  & $\cdots$ & $\cdots$ & $\cdots$ & $\cdots$ & $\cdots$ & $\cdots$\\
C & no &  & $3\times10^{-8}$ & $1.8$ & $10.0$ & $2.3$ & $10^{4}$ & $4.0$ & $2.2$ &  & $\cdots$ & $\cdots$ & $\cdots$ & $\cdots$ & $\cdots$ & $\cdots$\\
D & yes &  & $3\times10^{-8}$ & $5.8$ & $14.0$ & $2.3$ & $10^{4}$ & $4.0$ & $2.2$ &  & $\cdots$ & $\cdots$ & $\cdots$ & $\cdots$ & $\cdots$ & $\cdots$\\
E & yes &  & $4\times10^{-8}$ & $3.8$ & $12.0$ & $0.2$ & $10^{4}$ & $2.0$ & $2.2$ &  & $\cdots$ & $\cdots$ & $\cdots$ & $\cdots$ & $\cdots$ & $\cdots$\\
F & yes &  & $\cdots$ & $\cdots$ & $\cdots$ & $\cdots$ & $\cdots$ & $\cdots$ & $\cdots$ &  & $6\times10^{-8}$ & $35^{\circ}$ & $2.4$ & $10^{4}$ & $500$ & $1.2$\\
G & no &  & $\cdots$ & $\cdots$ & $\cdots$ & $\cdots$ & $\cdots$ & $\cdots$ & $\cdots$ &  & $6\times10^{-8}$ & $35^{\circ}$ & $2.4$ & $10^{4}$ & $500$ & $1.2$\\
\hline 
\end{tabular} 
Note: ($\dagger$) MA indicates if a model includes a magnetospheric
accretion. If indicated by `yes', the parameters in Table~\ref{tab:stellar_parameters}
are used for the magnetosphere. ({*})~Model~A consists of the magnetospheric
accretion flows only.
\end{table*}
%=========================================

%************************************************************
%************************************************************

\subsubsection{Bipolar outflow}
\label{subsub:SW-config}

As an alternative flow geometry, we also consider a
bipolar outflow which arises in the polar regions near the star. Possible scenarios in which this type of outflows could occur
are: (1)~an accretion-powered stellar
wind launched from open magnetic field lines anchored to the stellar surface
(e.g.~\citealt{decampli81,strafella98,bouret00,matt05,cranmer08,cranmer09}),
or (2)~a relatively fast magnetically driven polar outflow which
appears in between the slower conically-shaped winds 
launched from the disk-magnetosphere interaction regions
(e.g.~\citealt{romanova09,lii12}). 

The bipolar outflow is approximated as an outflow consisting of narrow cones with
a half-opening angle $\theta_{\mathrm{bp}}$. Here, we assume that the
flow propagates only in the radial direction, and that its velocity is described by
the classical beta-velocity law (cf.~\citealt{castor79}):
\begin{equation}
 v_{r}\left(r\right)=v_{0}+\left(v_{\infty}-v_{0}\right)\left(1-\frac{R_{0}}{r}\right)^{\beta_{\mathrm{bp}}}\,,
 \label{eq:beta-velocity-law}
\end{equation}
where $v_{\infty}$ and $v_{0}$ are the terminal velocity and the
velocity of the outflow at the base ($r=R_{0}$). Here, we set
$v_{0}=10\,\mathrm{km\, s^{-1}}$ which roughly corresponds to the
thermal velocity of gas around $10^{4}$~K.

Assuming that the mass-loss rate of
the outflow is $\dot{M}_{\mathrm{bp}}$ and
following the mass-flux conservation in the flows, the density $\rho_{\mathrm{pb}}$ of the outflow can be written as:
\begin{equation}
\rho_{\mathrm{bp}}\left(r\right)=\frac{\dot{M}_{\mathrm{bp}}}
   {4\pi
     r^{2}v_{r}\left(r\right)\left(1-\cos\theta_{\mathrm{bp}}\right)}\,.
  \label{eq:sw-density}
\end{equation}
The temperature of the bipolar outflow ($T_{\mathrm{bp}}$) is assumed
isothermal as in~\citet{kurosawa11}. To avoid an overlapping of the
bipolar outflow with the accretion funnels, the base of the bipolar outflow
($R_{0}$) is set approximately at the outer radius of the
magnetosphere ($R_{\mathrm{mo}}$) (cf.~Fig.~\ref{fig:model-config}).
The rotational velocity of the flow is assigned such that the specific angular
momentum of gas is conserved along the stream lines, and its initial
value is constrained by the rotation speed of the stellar surface. 
A similar bipolar outflow model was recently applied to the pre-FUor star,
V1331~Cyg \citep{petrov14}. 

The bipolar outflow introduced here is rather a simple
  kinematic model which does not involve a physical mechanism for the
  wind acceleration. %, and somewhat less sophisticated than the disc
  %wind model introduced in Section~\ref{subsub:DW-config}. 
  However,
  since the outflow geometry here is drastically different from that
  of a disc wind model, it is useful to consider this model to
  qualitatively check if the outflow in VV~Ser arises mainly from the
  polar regions or the accretion disc. 

\subsection{Modelling line profiles}

In this section, we will apply the models described above (i.e.
magnetospheric accretion, disc wind and bipolar  outflow) to model the \hi\ line profiles obtained from the LBT
observations.  Due to the complex line profiles and line variability
observed in these data, we opted to model, as a first approximation,
the average line profiles of the \brg, \pab\ and \pad\ lines during
the three epochs. These data offer a higher spectral resolution
(R$\sim$7000) than the MR-AMBER data (R$\sim$1500).  The origin of the
line variability will be discussed in Section\,\ref{sect:discussion}.

\subsubsection{Adopted stellar and disc parameters}
\label{subsub:adopted-param}

%************************************************************
\begin{figure}
\begin{center}
\includegraphics[width=0.38\textwidth]{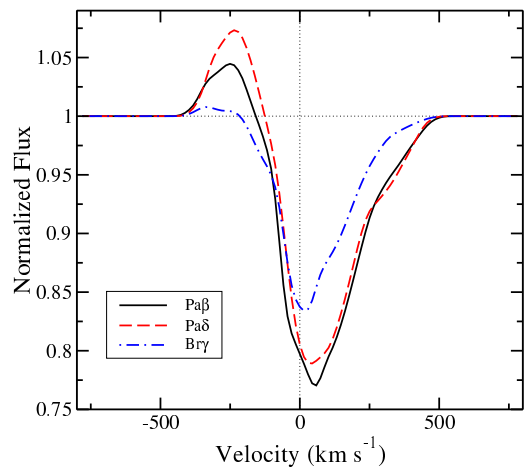}
\end{center}
\caption{Pa$\beta$ (solid), Pa$\delta$ (dash) and Br$\gamma$
  (dash-dot) model line profiles for VV~Ser computed (at
  $i=70^{\circ})$ with a small magnetosphere with
  $R_{\mathrm{mi}}=1.3\,R_{*}$ and $R_{\mathrm{mo}}=1.8\,R_{*}$ and
  the stellar rotation period of $P_{*}=0.84$\,d.  Other model
  parameters are summarised in Table~\ref{tab:stellar_parameters}.  The
  lines are mainly in absorption except for a very small emission
  component that arises on the blue side of the profiles. The model
  profiles clearly disagree with the observed profiles which display
  instead a relatively strong emission peak around the line centres
  (Fig.\,\ref{fig:LBT-obs}). For clarity, the continuum
  emission from the ring emission around the dust sublimation radius
  and photospheric absorption are excluded in these models.}
\label{fig:MA-demo}
\end{figure}
%************************************************************

The basic stellar parameters adopted for modelling the VV\,Ser spectra from the  LBT
are listed in Table\,\ref{tab:stellar_parameters}.
%are: $M_{*}=4.0\,\mathrm{M_{\odot}}$,
%$R_{*}=2.6\,\mathrm{R_{\odot}}$ and $T_{\mathrm{eff}}=14000$~K, as
%found in Sections~XXX and XXX.  
The atmosphere model of \citet{kurucz79} with
$T_{\mathrm{eff}}=14000$~K and $\log g=4.0$ (cgs) is used as the source
of the stellar continuum in the line radiative transfer models. A disk inclination angle $i$ of $\sim$70\degr\ is assumed \citep{pontoppidan07, eisner14}.
The speed at which the star rotates at the equator ($v_{\mathrm{eq}}$) was estimated assuming that the disc/ring normal vector coincides with the rotation axis of VV\,Ser, and that $v_{\mathrm{eq}}$ is the same as $v$ in $v\sin i$. 
In the case of VV\,Ser,  $v\sin i=142\,\mathrm{km\,
  s^{-1}}$ (\citealt{vieira03}), and thus $v_\mathrm{eq}=151\,\mathrm{km\, s^{-1}}$ (approximately $0.3$ of the breakup
velocity). 
%
%Assuming that the disc/ring normal vector coincides with the
%rotation axis of VV\,Ser and that the speed at which the star rotates at the
%equator ($v_{\mathrm{eq}}$) is the same as $v$ in $v\sin i$, we infer
%$v_{eq}=151\,\mathrm{km\, s^{-1}}$ (approximately $0.3$ of the breakup
%velocity) using an observed value of $v\sin i=142\,\mathrm{km\,
%  s^{-1}}$ (\citealt{vieira03}). 
Consequently, the rotation period
of VV~Ser is estimated to be $P_{*}=2\pi
R_{*}/v_{\mathrm{eq}}=0.86$~d. This value is in agreement with the one estimated by
\citet{hubrig09}. 
%We adopt the distance to VV~Ser as $d=415$~pc, which is the
%astrometric distance to the Serpens star forming cloud measured by
%Very Long Base Array \citep{dzib10}. 
Finally, a mass-accretion rate of 3.3$\times$10$^{-7}$\,\msyr is assumed.
%\footnote{This is slightly higher than the value found in Section~\ref{sub:acc-prop}; however, overall
%    behaviours of our models do not change
%     by adopting this value.}
%estimated from the line luminosity of Br$\gamma$ in Section XXX,
%i.e. $\dot{M}_{\mathrm{a}}=3.3\times10^{-7}\,\mathrm{M_{\odot}\,yr^{-1}}$.

To model the dust disc of VV~Ser, 
a uniform geometrical ring model is used to fit the AMBER continuum
visibility data in Section~\ref{sect:interferometry}. The inner radius 
of the ring is estimated to be 2.35~mas which corresponds to 0.97~au at
$d=415$~pc. We adopt the same ring geometry and dimensions as in the line
profile calculations. We also adopt the position angle (PA) and the ring
inclination angle ($i$) from our earlier analysis (Table\,\ref{tab:geometric_model}),
i.e. PA=\,66\fdg7 and $i$=70\degr, respectively. The ring is
assumed to emit as a blackbody at a temperature
$T_{\mathrm{ring}}=1400$~K.

\subsubsection{Line profiles from magnetospheric accretion models}
\label{subsub:MA-model-LBT}

In this section, we briefly discuss the effect of magnetospheric accretion
(see Section~\ref{subsub:MA-config}) on the formation of the NIR
atomic hydrogen emission lines observed in VV\,Ser. 
%We mainly focus on the models
%parameters used for VV Ser. 
Using the stellar parameters listed in Table\,\ref{tab:stellar_parameters}, the corotation radius ($R_{\mathrm{cr}}$) of VV~Ser becomes
$R_{\mathrm{cr}}=\left(GM_{*}\right)^{1/3}\left(P_{*}/2\pi\right)^{2/3}=2.4\,
R_{*}$.  We assume that the extent of the magnetosphere is slightly
smaller than $R_{\mathrm{cr}}$, and set the inner and outer radii of
the magnetospheric accretion funnel to be $R_{\mathrm{mi}}=1.3\,
R_{*}$ and $R_{\mathrm{mo}}=1.8\, R_{*}$.  These radii (in the units
of the stellar radius) are considerably smaller than those of
CTTSs (e.g.~\citealt{koenigl91};
\citealt{muzerolle04}). Consequently, the volume of the gas emitting
region/accretion funnel becomes very small. Hence, the hydrogen line 
emission strengths are expected to be also very small. The weakness of
emission lines in the small and fast-rotating magnetosphere was first
pointed out by \cite{muzerolle04}, in modelling the Balmer lines of
Herbig Ae star UX~Ori (see their fig.~3).

Figure~\ref{fig:MA-demo} shows the Pa$\beta$, Pa$\delta$ and Br$\gamma$
model line profiles assuming that they are only formed in the magnetosphere.  As expected, the lines are rather
weak, and mainly in absorption except for a very small emission
component on the blue side of the profiles. Very similar line shapes
are also found in the rotating magnetosphere model of
\cite{muzerolle04} (their fig.~3). The model clearly disagrees with
the observed line profiles (Fig.\,\ref{fig:LBT-obs}) which display
 strong emission peaks around the line centres. In particular,
the observed Pa$\beta$ and Br$\gamma$ emission lines are roughly
symmetric around the line centres. 

Although not shown here, we have
computed models with various combinations of magnetospheric
temperature $T_{\mathrm{m}}$ and mass-accretion rates
($\dot{M}_{\mathrm{a}}$). The line shapes of those models are similar to those shown in Fig.\,\ref{fig:MA-demo}. 
In general, the symmetric, or almost symmetric, emission line
profiles are difficult to explain with a fast-rotating
magnetosphere model alone, especially at a high inclination angle, as
in VV~Ser. For this reason, it is most likely that an additional gas
flow component is involved in the formation of the hydrogen emission
lines.

It must be noted, however, that the results presented here may not apply to other Herbig
Ae/Be stars. To examine whether the results above are applicable to
Herbig Ae/Be stars in general, a further investigation is needed. For instance, one would need to explore much larger parameter spaces. However, this is beyond the scope of the present paper and shall be investigated in the future. 

\subsubsection{Line profiles from disc wind models }
\label{subsub:DW-model-LBT}

To overcome the weakness of the line emission in the models with a
magnetospheric accretion funnel (Fig.~\ref{fig:MA-demo}), an
additional emission volume from a disc wind (see Section~\ref{subsub:DW-config}) will be added to the magnetospheric contribution. 

\begin{figure}
\begin{center}
\includegraphics[width=70mm]{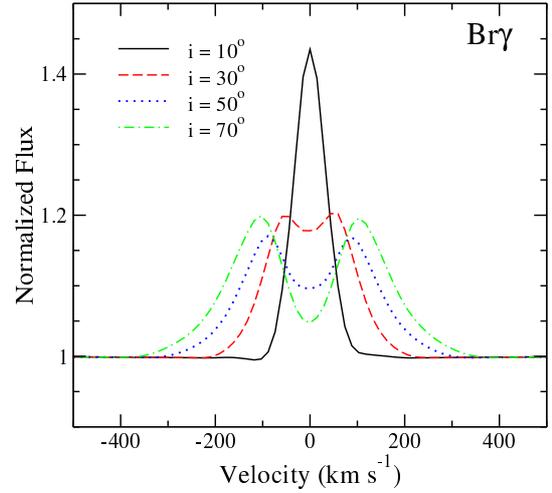}
\end{center}
\caption{The dependency of Br$\gamma$ on inclination angle ($i$) in
  Model~B which uses a combination of a disc wind and magnetosphere
  (Table~\ref{tab:model-param}).  The line profiles are double-peaked
  for the models with $i \simgreat 30^{\circ}$. Disc models in
  general have difficulties reproducing the singe-peak Br$\gamma$
  line profile seen in the LBT observations (Fig.\ref{fig:LBT-obs})
  for the models with mid to high inclination angles. 
  %N.B. the AMBER
  %observations (Fig.~\ref{fig:AMBER}) suggest that the inclination
  %angle of VV~Ser is $i\approx 70^{\circ}$. 
  }
\label{fig:DW-inc-effect}
\end{figure}

The left
panels in Fig.~\ref{fig:DW-compare-obs} show the model line profiles,
computed at $i=70^{\circ}$ (i.e. almost edge-on), resulting from the combination of a magnetospheric
accretion funnel and a disc wind (Model~B). The main model parameters
are summarised in Table~\ref{tab:model-param}. The ratio of the wind
mass-loss rate to the mass-accretion rate
($\dot{M}_{\mathrm{dw}}/\dot{M}_{\mathrm{a}}$) is $\sim0.09$. This
model produces similar line strengths and widths as seen in the
observed Pa$\beta$, Pa$\delta$ and Br$\gamma$ lines of the LBT observations. However,
the model profiles clearly disagree with the observed profiles in
their shapes: the former are double-peaked, but the latter are
single-peaked. 
%This is because on the one hand, the line emission mainly originates from
%the disc wind and the contribution of the magnetosphere to the line profile is very small (see Section\,\ref{subsub:MA-model-LBT}). An example of the contribution of each of the model components to the line profile is 
%shown in Fig.~\ref{fig:DW-compare-obs}. This figure shows a comparison of the line profiles produced by the
%magnetosphere plus disc wind model (Model~B), and those generated by the single contribution of the disc wind model
%(Model~C), and the magnetosphere model (Model~A). The figure clearly illustrate that the overall shape of the HI lines in model\,B (magnetosphere + disc wind) is dominated by the emission from the disc wind.
%This, on the other hand, has a strong impact on the shape of the modelled line profiles that become double-peaked. 
This is because in our disc wind model 
the \hi\ NIR lines are mainly formed near the
base of the disc wind. In this region the Keplerian rotation of the wind dominates over the radial motion. Therefore, double-peaked profiles naturally
occur when the disc wind is viewed from horizontal directions, that is, when the system has a mid to high inclination angle ($i$). This is demonstrated in Fig.~\ref{fig:DW-inc-effect}, which shows how the
Br$\gamma$ line profile depends on the inclination angle. In particular, this figure shows that the modelled Br$\gamma$ line profile becomes double peaked for $i\simgreat30^{\circ}$.  Note that in our disc wind model, the Keplerian velocities are about
$400\,\mathrm{km\, s^{-1}}$ and $170\,\mathrm{km\, s^{-1}}$ at the
inner and outer radii of the wind launching regions, that is, $R=1.8\,\mathrm{R_{\odot}}$ and $10\,\mathrm{R_{\odot}}$, 
respectively. Moreover, the line emission mainly originates from
the disc wind and the contribution of the magnetosphere to the line profile is very small (see Section\,\ref{subsub:MA-model-LBT}). This is illustrated in the right panels of
Fig.~\ref{fig:DW-compare-obs}, where a comparison of the line profiles produced by the
magnetosphere plus disc wind model (Model~B), with those generated by the single contributions of the disc wind model
(Model~C), and the magnetosphere model (Model~A) is shown. Therefore, in model\,B (magnetosphere + disc wind), the overall shape of the \hi\ lines is dominated by the emission from the disc wind.

Although not shown here, we have explored various combinations of the
disc wind model parameters (i.e.~$\dot{M_{\mathrm{dw}}}$,
$R_{\mathrm{wi}}$, $R_{\mathrm{wo}}$, $D$, $T_{\mathrm{dw}}$
$\beta_{\mathrm{dw}}$ and $p$ as in
Section~\ref{subsub:DW-config}; see also
Fig\@.~\ref{fig:model-config}) which control the characteristics of
the wind. 
 Examples of the dependency of the Br$\gamma$ profile on a few
selected models parameters (the wind launching radii $R_{\mathrm{wi}}$ and
$R_{\mathrm{wo}}$; the power-law index of local mass-loss rate per
unit area $p$) are given in Appendix~\ref{sec:appendix-dw-param}.
Even though the line strengths, line widths and the peak separation
widths vary with different combinations of the disc wind parameters,
the double peaks are always visible for a mid to high inclination angle, as is the case for VV~Ser
(assuming $i\approx70^{\circ}$). Hence, the disc wind is unlikely a major
contributor to the single-peaked emission lines (Pa$\beta$, Pa$\delta$
and Br$\gamma$) seen in VV Ser. 
However, if the inclination is $\sim$70\degr\ or larger, we cannot exclude the possibility
that the innermost region of the disc wind is obscured by the outer region
of a flared disc, and thus that another flux contribution is dominant.
For these reasons, we seek an alternative flow 
%
%Consequently, we seek an alternative flow
component which can robustly reproduce a rather symmetric and
single-peak Br$\gamma$ emission line as seen in VV\,Ser and many other
Herbig Ae/Be stars (e.g.~\citealt{rebeca06};
\citealt{brittain07}).  Note that, for a relatively low inclination
angle system (nearly face-on), it may be still possible to reproduce a
single peaked Br$\gamma$ by a disc wind model
(e.g.~Fig.~\ref{fig:DW-inc-effect}; see also \citealt{weigelt11, rebeca15, ale15}).

%************************************************************
%************************************************************

\subsubsection{Line profiles from bipolar outflow models }
\label{subsub:SW-model-LBT}

Another possible way to obtain a single-peak emission line profile in a nearly edge-on system is to
include a bipolar outflow in the model. 
%In addition, the system
%must be viewed at a relatively large inclination angle. 
%In other
%words, the inclination angle $i$ must be larger than the half-opening
%angle of the bipolar stellar wind $\theta_{\mathrm{bp}}$ (see
%Fig.~\ref{fig:model-config}) to avoid a strong wind absorption on the blue
%side of the line profile (e.g.~a P-Cygni profile).
In this case, the inclination angle $i$ must be larger than the half-opening
angle of the bipolar outflow $\theta_{\mathrm{bp}}$ (see
Fig.~\ref{fig:model-config}) to avoid a strong wind absorption on the blue
side of the line profile (i.e. a P-Cygni profile).

The left panels in Fig.~\ref{fig:SW-LBT} shows the model line
profiles computed with a combination of a magnetospheric accretion
funnel and a bipolar outflow (Model~F). The lines are computed at
$i=70^{\circ}$ assuming a wind half-opening angle $\theta_{\mathrm{bp}}=35^{\circ}$. The model parameters are
summarised in Table~\ref{tab:model-param}.  The ratio of the mass-loss
rate in the wind to the mass-accretion rate
($\dot{M}_{\mathrm{dw}}/\dot{M}_{\mathrm{a}}$) is about $0.18$.  The
figure clearly shows that the model is capable of producing
single-peak line profiles. Furthermore, the model produces line profiles very similar to the observed Pa$\beta$, Pa$\delta$ and Br$\gamma$ line profiles (in their line strengths and widths). 

As for the case of the magnetosphere-plus-disc wind model, the line emissions in this
model are mainly from the bipolar outflow. This is clearly illustrated in the right
panels of Fig.~\ref{fig:SW-LBT}, which compare the profiles from
Model~F (magnetosphere + bipolar outflow) with those computed with only
the bipolar outflow (Model~G) and only the magnetosphere (Model~A). In this
particular model (Model~F), the magnetosphere contributes only
slightly to the lines, and the overall shapes of the lines are
determined by the emission from the bipolar outflow.

\begin{figure}
  \begin{center}
    \includegraphics[width=84mm]{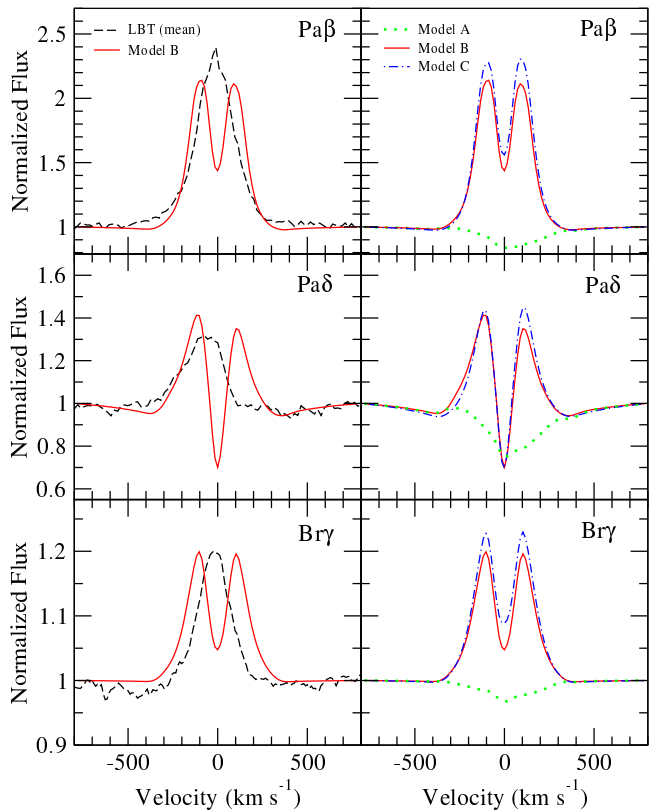}
  \end{center}

  \caption{Left Panels: a comparison of the mean line profiles
    obtained by LBT (dashed) with the hydrogen line
    profiles (Pa$\beta$, Pa$\delta$ and Br$\gamma$) computed using the
    models with the combination of a disc wind and magnetosphere
    (solid; Model~B in Table~\ref{tab:model-param}).  Right panels: a
    comparison of line profiles computed using the disc wind +
    magnetosphere (solid; Model~B) with those using (1) only the disc
    wind (dash-dot; Model~C) and (2)~only the magnetosphere (dot;
    Model~A).  All the profiles are computed at the inclination angle
    $i=70^{\circ}$. In general, models with disc wind produced
    double-peak profiles due to rotating motion of the disc wind. The
    double peaks are always prominent at a mid to a high inclination
    angle (cf.\,Fig.~\ref{fig:DW-inc-effect}). The right panels show
    that the emission contribution from the magnetosphere is very
    small in Model~B. }
  \label{fig:DW-compare-obs}
\end{figure}

The emission profiles of the Pa$\beta$ and Br$\gamma$ lines are mostly symmetric
around the line centres. However, the Pa$\delta$ line profile is slightly
asymmetric towards the blue side of the
line, that is, the blue-shifted emission is slightly stronger. 
This is due to the absorption by the magnetospheric accretion
funnel which is stronger on the red side of the line. A similar trend is 
also observed in the Pa$\delta$ line computed with the disc wind +
magnetosphere (Model~B) in Fig.~\ref{fig:DW-compare-obs}, which shows
stronger blue-shifted emission.

In summary, the bipolar outflow, when viewed from equatorial
directions, can form single-peak line profiles as the ones observed in the LBT spectra. 
The line strengths and shapes of the observed
Pa$\beta$, Pa$\delta$ and \brg\ line profiles are reproduced well
by Model~F (Fig.~\ref{fig:SW-LBT}; Table~\ref{tab:model-param}).

\subsection{Models for AMBER data}
\label{sub:models-AMBER}

To probe whether the  same bipolar outflow plus magnetosphere model described above can reproduce the \brg\ AMBER-VLTI spectro-interferometric observations (Section\,\ref{sect:interferometry}) in addition to the 
LBT spectra, the intensity maps of the line emitting regions were computed using the same radiative transfer model employed to produce the line profiles (see Section~\ref{sect:models}). 
In this way, the interferometric
quantities such as visibility, differential and closure phases can be
readily calculated from the model intensity maps.

%In this section, we examine whether the stellar wind + magnetosphere model that
%can reproduce the observed near-infrared hydrogen emission lines
%(Fig.~\ref{fig:SW-LBT}), can also account for the VLTI+AMBER
%spectro-interferometric data around Br$\gamma$ presented in
%Section\,\ref{sect:interferometry}. 
%Our radiative transfer model, described in
%Section~\ref{sec:models}, can produce not only line profiles, but also
%the intensity maps of line emitting regions. The interferometric
%quantities such as visibility, differential and closure phases can be
%readily calculated from the model intensity maps.

The results are summarised in Fig.~\ref{fig:DW-SW-AMBER}. The figure
compares the model Br$\gamma$ line profiles, visibility and
differential phases with those of LBT (average \brg\ line profile) and VLTI-AMBER
observations.  For purposes of comparison, the figure also includes
the models computed with a disc wind plus a magnetosphere in addition to
those with the bipolar outflow.  
Six different cases in Table~\ref{tab:model-param} are shown here: 
(1)~a disc wind with small launching radii (Model~B) computed at
$i=70^{\circ}$, (2)~a disc wind with larger launching radii (Model~D) computed at $i=70^{\circ}$,
(3)~a disc wind with intermediate launching radii and a small $p$ (the
index of the power-law in the mass-loss rate per unit area; see
Appendix~\ref{sec:appendix-dw-param} and \citealt{kurosawa11})
(Model~E) computed at $i=70^{\circ}$, (4)~a bipolar outflow model
(Model~F) computed at $i=65^{\circ}$, (5)~same as (4) but computed at
$i=70^{\circ}$, and (6)~same as (4) but with $i=75^{\circ}$.  
Note that model closure phases are not shown in this figure because no
closure-phase signal within the error of  $20^\circ$ is measured from 
our AMBER observations (see, Fig.\,\ref{fig:AMBER-obs}), and all model closure phases are smaller than 10\degr. 

In cases~(1)--(3) (disc wind models), the models fit the
interferometric observations reasonably well. However, the 
double-peak line profiles seen in those models do not match the
observations.  The separation of the double peaks are slightly reduced
by increasing the disc wind launching radii from case~(1) (Model~B) to
case~(2) (Model~D) (Table~\ref{tab:model-param}). A further
reduction of the peak separation is achieved by reducing the value of $p$ (the 
power-law for the mass-loss rate per unit area;
c.f.\,Appendix~\ref{sec:appendix-dw-param}; \citealt{kurosawa11}) 
from case~(2) (Model~D) to case~(3) (Model~E)
(Table~\ref{tab:model-param}). The double-peak feature is rather
persistent in the models with a disc wind
(see Section~\ref{subsub:DW-model-LBT}). On the other hand, the
bipolar outflow models (Model~F) with $i=65^{\circ}$ (case~4) and $75^{\circ}$
(case~6) reproduce reasonably well the observed line profiles, but the
visibility levels disagree with the AMBER observations. Only the model with
the bipolar outflow computed at $i=70^{\circ}$, case~(5), fits the
observed line profiles (here and also in Fig.~\ref{fig:SW-LBT}),
visibility and differential phase,
simultaneously.

In all the models, the visibility levels are slightly larger in the lines
(by $\sim0.06$) compared to those in the continuum. This agrees with the
fact that the line emitting regions are smaller than that of the
continuum emitting regions. In our models, the inner radius of the
K-band continuum emitting ring model is approximately 1.0~au
(Section\,\ref{sect:geometric_model}) which corresponds to
$\sim80\,\mathrm{R_{\odot}}$.  This is much larger than the wind
launching regions of both disc wind ($1.8$--$14\,\mathrm{R_{\odot}}$)
and bipolar outflow ($2.4\,\mathrm{R_{\odot}}$)
(Table~\ref{tab:model-param}). The increase in the visibility levels across the Br$\gamma$ emission line is not clearly seen in the AMBER observations due to the relatively low S/N and spectral resolution of the data. 
However, a weak increase of the visibility at the longest baseline (UT1--UT4) is observed  towards the line centre (see Fig.\,\ref{fig:AMBER-obs}). The small increase in the visibility in the line centre
($\sim0.06$) with respect to the continuum seems to be in agreement
with the data, considering the levels of uncertainty/noise.

The differential phases from the AMBER observation are $0^{\circ}$
within the error of $5^{\circ}$ across the Br$\gamma$ line
without any 
trend for an increase or a decrease. On the other hand, our models show
very small variations with their amplitude between $\sim1^{\circ}$ and
$\sim2^{\circ}$ across the line. These small
amplitudes are well below the noise levels of the observations.  This
suggests that the models do not disagree with the observed
differential phases, that is, the amplitude of the variation in the
differential phase across the line predicted by the model do not
exceed the noise levels in the observations. 
%The small differential
%phases displayed by the observation and models indicate that the
%physical separation between the line emitting regions moving at blueshifted and redshifted velocities 
%%positive radial velocities (moving away from the observer) and the region emitting at 
%%negative radial velocities (moving towards the observer) 
%is
%relatively small.

%*******************************************************************
%****************************************************************
In principle, the physical parameters of the line emitting wind
regions (the wind geometry and kinematics) can be further constrained
by performing detail model fitting of the observed visibility and
differential phases across the line. However, the high noise
levels in the data prevent us from performing such an investigation. 

In summary, the model with a bipolar outflow
 (Model~F) observed at an inclination angle of 
$i=70^{\circ}$ (the third column in Fig.~\ref{fig:DW-SW-AMBER}) gives
the best match to the observed visibility, differential phases and
the NIR hydrogen line profiles (Pa$\beta$, Pa$\delta$ and Br$\gamma$
as in Fig.~\ref{fig:SW-LBT}).

\section{Discussion}
\label{sect:discussion}

\subsection{Origin of the NIR \hi\ emission lines}
\label{sect:origin_HI}

In the previous sections, we investigated possible mechanisms responsible for the formation of the NIR \hi\ emission lines in the Herbig star VV Ser. With this goal in mind, we explored 
what should be expected from three different 
line emitting sources namely, a magnetosphere, a disc wind and a
bipolar outflow, and compare these results with
average \hi\ line profiles obtained with LUCIFER at LBT, and with
AMBER-MR spectro-interferometric observations.  
Our modelling suggests, that in the case of VV\,Ser, the contribution of a fast rotating magnetosphere to the \hi\ NIR emission line profiles is very small in comparison with the 
contribution of a disc wind or bipolar outflow. 
Similar results have been previously found in other Herbig AeBe stars by analysing their NIR line profiles alone or in combination with NIR spectro-interferometric  techniques \citep{weigelt11,larisa14,rebeca15, ale15}.
In contrast with these results, where a disc wind model could be used
to model the \brg\ line profile and the spectro-interferometric
observables, in the case of VV\,Ser only a bipolar outflow %(probably in combination with a weak emission from a magnetosphere) 
can  approximately reproduce the
averaged single-peaked line profiles of the \pab,
\pad, and \brg\ lines observed with the LBT, as well as, the
\brg\ spectro-interferometric observables. 
The main difference,  
between VV\,Ser and the results mentioned above is the high disc inclination angle of VV\,Ser ($i\sim$70\degr). As shown in Fig.\,\ref{fig:DW-inc-effect}, our disc wind model 
 can only reproduce single peaked line profiles when low inclination angles ($i\lesssim$30\degr) are considered. Similar results are also found in \cite{larisa14} for the \brg\ and H$\alpha$ lines.
This is because, the NIR \hi\ lines are formed at the base of the
disc wind, where the Keplerian rotation of the disc dominates over the
wind radial velocity.
Therefore, a disc wind can only generate a single peaked line profile at high inclination angles if the inner disc wind emitting region is obscured by an outer flared disc, or alternatively, if other components (such as a magnetosphere or a bipolar wind)
significantly contribute to the line profile. In contrast, 
%Therefore, only when 
%a bipolar outflow, launched
bipolar outflows are launched 
well above the  circumstellar disc 
(see, sketch in Fig.\,\ref{fig:model-config}), 
%is considered, 
and thus,
 single peaked line profiles are 
easily
obtained for all the NIR \hi\ observed lines. 

%************************************************************
\begin{figure}
\begin{center}
\includegraphics[width=84mm]{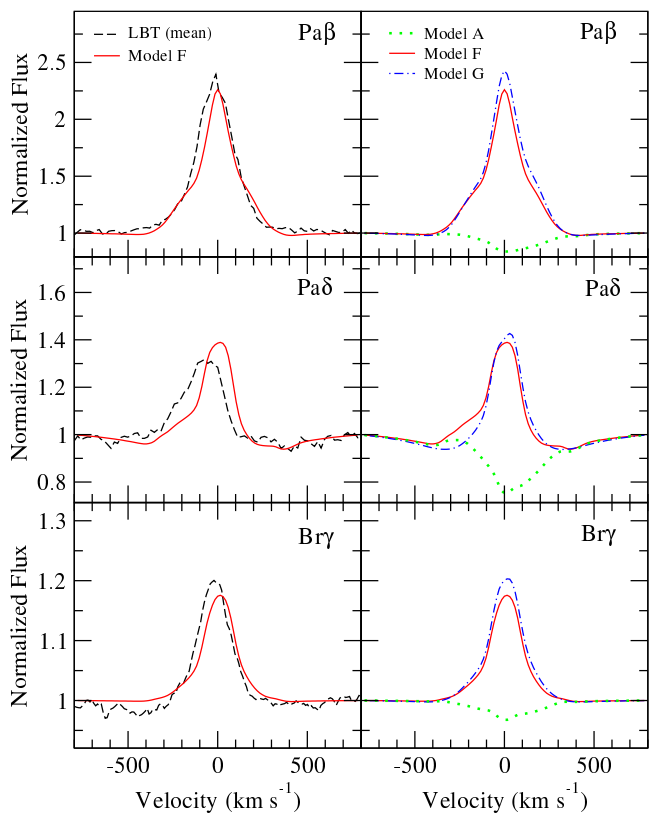} 
\end{center}
\caption{Left Panels: a comparison of the mean line profiles obtained
  by the LBT (dashed; see Section\,\ref{sect:spectroscopy} and Fig.\,\ref{fig:LBT-obs}) with the model hydrogen line
  profiles (Pa$\beta$, Pa$\delta$ and Br$\gamma$) computed using the
  models with the combination of a bipolar outflow and magnetosphere
  (solid; Model~F in Table~\ref{tab:model-param}). Right panels: a
  comparison of the line profiles computed using the
  bipolar outflow +   magnetosphere (solid; Model~F) with those using (1) only the
  bipolar outflow (dash-dot; Model~G) and (2)~only the magnetosphere (dot;
  Model~A). All the profiles are computed at the inclination angle
  $i=70^{\circ}$. The left panels show that Model~F can reasonably and
  simultaneously fit the observed Pa$\beta$, Pa$\delta$ and Br$\gamma$
  profiles (mean) obtained by the LBT.  The right panels
  show that the emission contribution from the magnetosphere is very
  small in Model~F.}
\label{fig:SW-LBT}
\end{figure}

%************************************************************

Although the bipolar outflow model works very well
  here, we would like to remind readers that the model is highly
  schematic, and the exact physical process which can produce this type
  of outflows is not well known. As mentioned earlier, one possible
  candidate is the accretion-powered stellar wind
  (e.g.~\citealt{decampli81,strafella98,bouret00,matt05,cranmer08,cranmer09});
  however, these models (except for \citealt{strafella98,bouret00})
  are mainly for winds in CTTSs which have stronger magnetic fields.  Interestingly,
  recent spectroscopic studies of Herbig Ae/Be stars by
  \citet{cauley14, cauley15} have shown that about 
  30~per~cent of their sample display P-Cygni profiles in
  HeI~10830\,\AA{} and H$\beta$, which is
  a good indication for stellar wind like outflows
  (e.g.~\citealt{edwards06,kwan07,kurosawa11}).

Our AMBER-MR spectro-interferometric observations of the \brg\ line support the results obtained from the pure modelling of the LBT spectroscopic observations. That is, 
%On the other hand, it should be mentioned here, that 
regardless of the accuracy on the line profile modelling, the wind
(disc or bipolar wind) component dominates the line emission, whereas the magnetosphere do not play a major role in the line profile.
Indeed, the AMBER-MR interferometric observations of the \brg\ line shows that even if the \brg\ line emission is more compact than the continuum emission, the \brg\ emission is nevertheless resolved. Therefore, the line emitting region 
is not mainly formed in the unresolved magnetospheric accretion
region. Furthermore, the interferometric observables (visibilities,
differential phases and closure phase) are best reproduced when a disc
wind or bipolar outflow is considered, 
the latter being the one that best reproduces the \brg\ single line profile.

\subsection{Line profile variability}
\label{sect:line_variability}

As shown in Fig.\,\ref{fig:variance_profile}, a significant fraction of the variation in the observed lines occurs in their red wings. Therefore, variations in the contribution
of the magnetosphere to the total \hi\ line emission could be the cause
of the NIR \hi\ line variability observed in VV\,Ser (see,
Fig.\,\ref{fig:LBT-obs}). For instance, a periodic and
  a non-periodic line variability can be caused by rotating non-axisymmetric
  magnetospheric accretion flows in stable and unstable regimes,
  respectively (e.g.~\citealt{kurosawa13}). Similarly, non-steady accretion or
  an enhanced
  mass-accretion rate could modify the redshifted component of the 
line profiles, introducing strong redshifted absorption components to the line profiles (see Fig.\,\ref{fig:MA-demo}). This could be the cause of the redshifted adsorptions in the \pad, \pab\ and \brg\ line profiles observed in February 2012 (see Fig.\,\ref{fig:LBT-obs}).  
On the other hand, the complex \pad\ line profile observed in April 2012 could be explained by an increase from a disc wind contribution to the line profile. A disc wind could also reproduce the double peaked \ha\ line profiles observed in VV\,Ser (see e.g., \citealt{mendigutia11}). This might indicate that lines such as \ha\ are optically thicker than the NIR \hi\ emission lines. 
Furthermore, they are formed in regions located farther away from the disc plane and observed through the wind down to the central source.
In this context, the complex line profiles and their variability could
be explained through the coexistence of at least three different
mechanisms, namely magnetospheric accretion, a disc wind and a
bipolar outflow, the intensity and contribution to the 
total line profile of which vary with time.
This would indicate that accretion, and thus, the ejection processes do not occur in a steady-state regime. 

Non steady accretion has been previously used to explain the short term spectral and photometric variability of the CTTS AA\,Tau \citep{bouvier03}. In this case, the optical line profile and photometric variations were explained by the presence of a non-axisymmetric warp 
in the inner disc caused by the interaction of the disc and an
inclined stellar dipole (e.g.~\citealt{romanova03,romanova13}). The variability is then related to the perturbation of the stellar magnetic field at the disc inner edge. In this regard, an increase in brightness, and the subsequent
line variability, is due to a severe reduction of the accretion flow onto the star, leading to a temporary disappearance of the disc inner warp.
However, it is not clear how such a mechanism could work in VV\,Ser. Whereas CTTSs are slow rotators  as a consequence of strong magnetic fields producing a magnetic braking, Herbig AeBe stars are fast rotators with no clear 
evidence of a {developed and organised  magnetic field configuration \citep[e.g.,][and references therein]{hubrig11}. Although the measured variations in the EW of the \hi\ lines could be interpreted as changes in the continuum level (i.e. magnitude) of the source, they could also represented a decrease in the 
line flux. Therefore, without simultaneous photometry at both optical and NIR wavelengths it is difficult to test whether VV\,Ser's variability is due to inner disc instabilities caused by enhanced/depressed accretion. Further observations, spanning a 
wider wavelength and temporal range, and accompanied by simultaneous optical and NIR photometry are needed in order to disentangle the inner disc mechanism producing the line variability.  

\section{Summary and conclusions}
\label{sect:conclusions}

%************************************************************
\begin{figure*}
\begin{center}
\includegraphics[width=175mm]{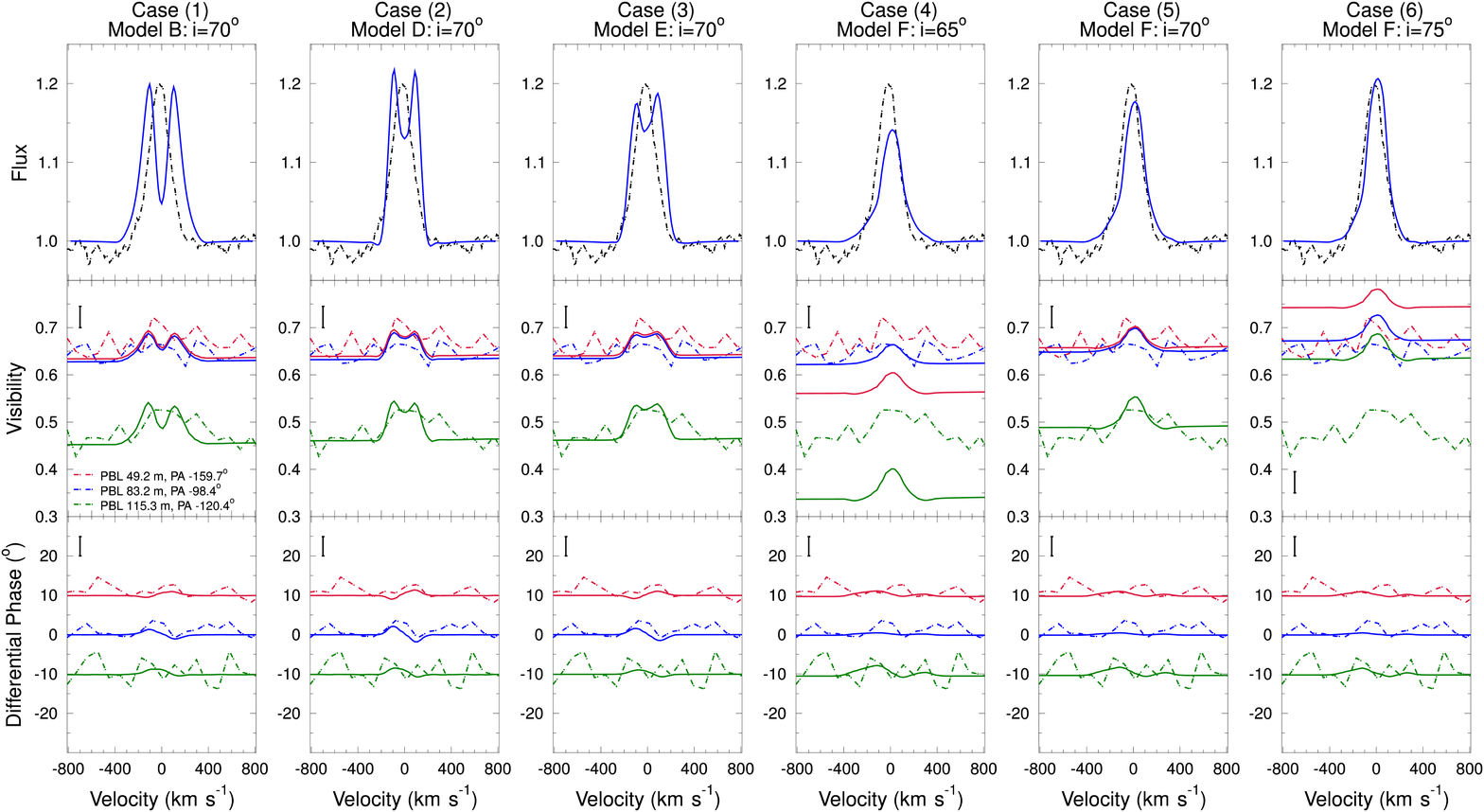}
\end{center}
\caption{Comparisons of the LBT and AMBER observations of Br$\gamma$
  with our models (six different cases): 
  (1)~Model~B with $i=70^{\circ}$ --- a disc wind with
  small launching radii (first column),
  (2)~Model~D with $i=70^{\circ}$ --- a disc wind with lager launching
  radii (second column), 
  (3)~Model~E with $i=70^{\circ}$ --- a disc wind with intermediate
  launching radii and a small $p$ (third column), 
  (4)~Model~F with
  $i=65^{\circ}$ --- a bipolar outflow
 model (fourth column),
  (5)~Model~F with
  $i=70^{\circ}$ --- a bipolar outflow
 model (fifth column), and 
  (6)~Model~F with
  $i=75^{\circ}$ --- a bipolar outflow
 model (sixth column).  See
  Table~\ref{tab:model-param} for their model parameters. 
  The top panels show the comparisons of the mean
  Br$\gamma$ profile from the LBT (dash-dot) with models (solid). The
  middle and bottom panels show the comparisons of our models (solid)
  and the visibility and differential phases obtained by AMBER
  (Section~\ref{sect:interferometry}) with three baselines: UT1--UT2
  ($\mathrm{PBL}=49.20$\,m, $\mathrm{PA}=-159.67^{\circ}$), UT2--UT4
  ($\mathrm{PBL}=83.25$\,m, $\mathrm{PA}=-98.40^{\circ}$) and UT1--UT4
  ($\mathrm{PBL}=115.28$\,m, $\mathrm{PA}=-120.38^{\circ}$). For
  clarity, the differential phases of UT2--UT4 and UT1--UT4 baselines
  are vertically shifted by $+10^{\circ}$ and $-10^{\circ}$,
  respectively.  Model~F with $i=70^{\circ}$ (fifth column)
  agrees reasonable well with the observed Br$\gamma$ profile, visibility
  and differential phase. The same model reproduces the observed
  Pa$\beta$ and Pa$\delta$ lines also well (Fig.~\ref{fig:SW-LBT}).
  The disc wind models (Models~B,D and E with $i=70^{\circ}$)
  fit the interferometric observations, but their line profile
  shapes (double peaked) do not match the observation. Although the
  bipolar outflow
  models (Model~F) with $i=65^{\circ}$ and $75^{\circ}$ reproduce reasonably well
  the observed line profile, their visibility levels do not match with
  the observation.}
\label{fig:DW-SW-AMBER}
\end{figure*}
%************************************************************

We have shown LBT-LUCIFER medium resolution z-, J-, and K-band spectroscopic observations covering the \pad, \pab, and \brg\ lines of the Herbig star VV\,Ser at three different epochs spanning over three months. 
Additionally, MR VLTI-AMBER interferometric observations of the \brg\ line in this source are 
presented as well. 

The spectroscopic observations show relatively strong line variability in all the \hi\ NIR emission lines. The line profile variability is more extreme in the 
Paschen lines probably because the line emitting volume of these lines is larger than that of the \brg\ line. 
%in a region that is higher above the disc plane 
%than the other lines.
Therefore, an observer would receive emission originating from different circumstellar regions (such as a wind and magnetospheric region) along the line of sight.
To investigate the physical mechanisms that are responsible for the \hi\ line
emission, we have a applied a radiative transfer model consisting of a
magnetospheric accretion region, a disc wind and a
bipolar outflow. The best model able to reproduce the 
roughly symmetric and single peaked averaged line
profiles of the \hi\ lines for the three epochs is a bipolar outflow model. However, it should be noted that the bipolar outflow model presented in Section~\ref{subsub:SW-config} is highly
  schematic, and  the exact physical process which can produce this type
  of outflows is not well known.
The contribution to the line profiles of the magnetospheric accretion region is small in comparison with that of any of our wind models. Therefore, the 
NIR \hi\ emission region is probably dominated by wind emission.
This is supported by our spectro-interferometric observations of the \brg\ line. Our results show a \brg\ emitting region smaller than the continuum emitting region. However, the \brg\ line emission is only resolved at the 
longest baseline. This indicates that the emission is compact but nevertheless spatially resolved (i.e. consistent with wind emission).

%that most of the emission is extended, and thus does not come from an unresolved magnetospheric accretion region. 

On the other hand, episodic enhancement accretion could explain the redshifted absorption components observed in some of our data. These periods could be followed by an increase of disc-wind activity. Although it is difficult to produce single peaked
line profiles from a disc wind model alone, some of the line profiles observed in VV\,Ser, such as the \ha\ line profile observed by \cite{mendigutia11} and the triple-peaked \pad\ line profile observed in April 2012 could be explained by the combined 
effect of a double peaked disc wind profile (see
Fig.\,\ref{fig:DW-compare-obs}) plus the single-peak profile obtained
in a bipolar outflow (see Fig.\,\ref{fig:SW-LBT}). 
Indeed, the AMBER visibilities, differential phases and closure phases
can be approximately reproduced by a disc 
wind or bipolar outflow, although, only a bipolar outflow (i.e. gas emitted close to the polar regions) can reproduce
the single peaked \brg\ line profile observed in  VV\,Ser (see Fig.\,\ref{fig:DW-SW-AMBER}).

In summary, our results indicate a very complex structure for the inner disc region of VV\,Ser, in where several mechanisms
(i.e. magnetospheric accretion, disc wind and bipolar outflow) could coexist in a non steady-state regime, given rise to strong line variability
and increasing the complexity of our understanding of the innermost gaseous disc region.

% --- Acknowledgments ---------------------- 
\section*{Acknowledgments}

We thank the anonymous referee for his/her useful comments and suggestions.
This research has made use of the Jean-Marie Mariotti Center
\texttt{Aspro} service \footnote{Available at http://www.jmmc.fr/aspro}. 
We thank Dmitry Shakhovskoy for the photometric follow-up of VV Ser.
R.G.L and A.C.G. were supported by Science Foundation of
Ireland, grant 13/ERC/I2907. A.K. acknowledges support from an STFC Rutherford Grant (ST/K003445/1).  L.V.T. was
partially supported by the Russian Foundation for Basic Research (Project 15-02-05399). V.P.G was supported by grant RFBR (project 15-02-09191).

% --- BibTex Bibliography ---------------------- 
\bibliographystyle{mnras}
\bibliography{references}

%---------------------------------------------------
\appendix

\section{Dependency of line profiles on selected disc wind parameters}
\label{sec:appendix-dw-param}

In Section~\ref{subsub:DW-model-LBT}, we found that the double peaks
in the emission line profiles of Pa$\beta$, Pa$\delta$ and Br$\gamma$
are typical features in the disc wind models in which the base of the
wind is rotating at nearly the Keplerian velocity of the accretion
disc.  This feature is especially persistent at high
inclination angles as in the case of VV~Ser
(Fig.~\ref{fig:DW-compare-obs}), and is inconsistent with the
single-peak appearance of the observed Pa$\beta$, Pa$\delta$ and
Br$\gamma$ lines (e.g.\,Fig.~\ref{fig:DW-compare-obs}). 

Here, we examine the dependency of the model line profiles on the disc
wind parameters which could potentially give rise to a single-peaked profile. We attempt this by reducing the separation of
the double peaks as much as possible, i.e. by trying to merge the
double peak into one. There are at least two possible ways of reducing the
peak separations. The first method is to increase the wind launching
radii ($R_{\mathrm{wi}}$ and $R_{\mathrm{wo}}$ as in
Fig.~\ref{fig:model-config}). By increasing the launching radius, the
rotational speed in the disc wind will be reduced. Consequently, the
separation of the double peaks would also be decreased.
Figure~\ref{fig:DW-Rwi-Rwo-effect} shows examples of Br$\gamma$ line
profiles computed with three different combinations of wind launching
radii. The figure shows that the separation of the double peaks
slightly decreases as the wind launching radii increase. The emission
in the line wings deceases significantly as we increase the wind
launching radii because the disc wind is launched with lower rotation
velocity at larger radii.  If we further increase the launching radii,
the separation of the double peak will decrease slightly, but the wing
emission will decrease significantly. Consequently, the line shape will be
inconsistent with the observations. In fact, in the line wings of the model
with the largest launching radii ($R_{\mathrm{wi}}=5.8\,R_{*}$ and
$R_{\mathrm{wo}}=14\,R_{*}$), the effect of the magnetospheric absorption
(near the velocity $\pm 200\, \kmps$ ) is visible 
due to the lack of wing emission in this model. The appearance of the
photospheric absorption in the line wing will be more pronounced as we
further increase the launching radii; hence, it would be
inconsistent with the observations.

The second method of decreasing the separation of the double peaks is
to reduce the power-law index ($p$) in the local mass-loss rate per
unit area, i.e.\, $\dot{m}(w) \propto w^{-p}$ where $w$ is the
distance from the star on the equatorial plane (e.g.\,
\citealt{kurosawa11}). A relatively large $p$ 
value leads to the wind mass loss being concentrated near the inner wind
launching radius ($R_{\mathrm{wi}}$). By decreasing the $p$ value, the
relative amount of mass loss in the outer part of the disc wind region
will increase. This leads to a reduction of the gas rotating at
relatively high speeds. Consequently, the separation of the double
peaks would be also decreased.  Figure~\ref{fig:DW-p-index--effect} shows
examples of Br$\gamma$ line profiles computed with four different
values of $p$. In all these models, we adopted the largest disc wind
launching radii used in Fig.~\ref{fig:DW-Rwi-Rwo-effect},
i.e.\,$R_{\mathrm{wi}}=5.8\,R_{*}$ and $R_{\mathrm{wo}}=14\,R_{*}$.
The figure shows that the peak separation decreases when $p$
decreases, as expected. However, even with the lowest possible value $p=0$
(i.e. $\dot{m}$ is uniform over the wind launching region), the double
peaks in the model Br$\gamma$ are still present.

%************************************************************
\begin{figure}
\begin{center}
\includegraphics[width=0.4\textwidth]{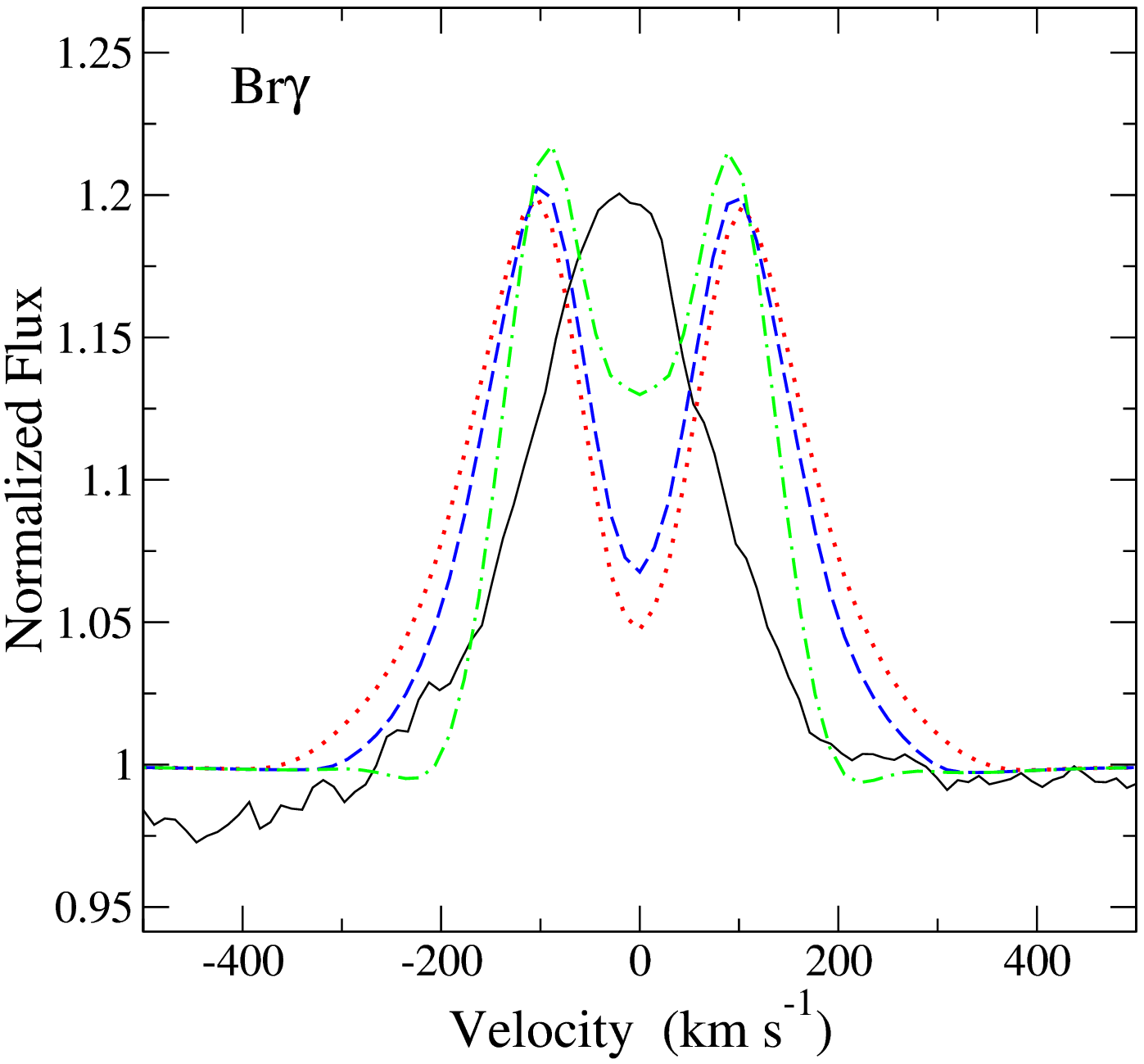}
\end{center}
\caption{The dependency of Br$\gamma$ on the disc wind launching
  radii.  The mean line profile of Br$\gamma$ LBT observations
  (Section~\ref{sect:spectroscopy}) is compared with models computed with three different
  combinations of the wind launching radii, $(R_\mathrm{wi},R_\mathrm{wo}) =
  (1.8\,R_{*},10\,R_{*})$ (dot),  
  $(2.8\,R_{*},12\,R_{*})$ (dash) and $(5.8\,R_{*},14\,R_{*})$
  (dash-dot), are shown. 
  The separation of the double peaks and the line widths of
  the model line profiles decrease as the launching radii increases. 
  All other model parameters used here are the same as in Model~B
  (Table~\ref{tab:model-param}). The line profiles are computed at the
  inclination angle $i=70^{\circ}$.
 }
\label{fig:DW-Rwi-Rwo-effect}
\end{figure}

%************************************************************

%************************************************************
\begin{figure}
\begin{center}
\includegraphics[width=0.4\textwidth]{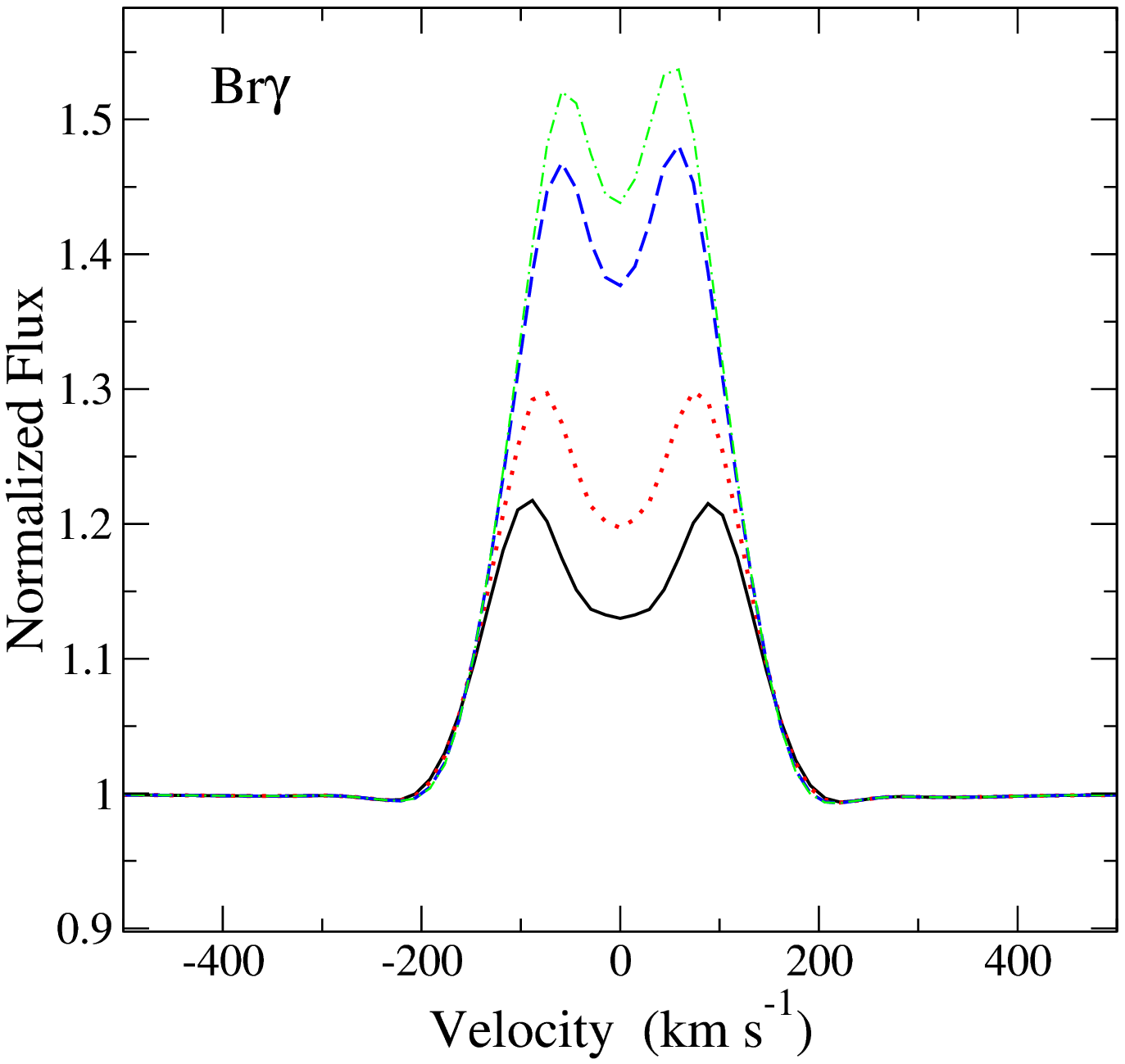}
\end{center}
\caption{The dependency of Br$\gamma$ on the power-law index ($p$) of
 the local mass-loss rate per unit area, $\dot{m}$ (as in
  Section~\ref{subsub:DW-config}).  The model line profiles are
  computed with $p=2.3$ (solid), $1.4$ (dot), $0.2$ (dash) and $0.0$
  (dash-dot).  The separation of the double peaks decreases as $p$
  decreases.  All other model parameters used here are same as in
  Model~D (Table~\ref{tab:model-param}). The line profiles are
  computed at the inclination angle $i=70^{\circ}$.
}
\label{fig:DW-p-index--effect}
\end{figure}

%************************************************************

\bsp

\label{lastpage}

\end{document}